\documentclass[conference]{IEEEtran}
\IEEEoverridecommandlockouts
\usepackage{cite}
\usepackage{amsmath,amssymb,amsfonts}
\usepackage{algorithm}
\usepackage{algpseudocode}
\usepackage{graphicx}
\usepackage{multirow}
\usepackage{textcomp}
\usepackage{xcolor}
\usepackage{subfigure}
\usepackage{subcaption}
\usepackage[top=1in,left=0.95in,right=0.95in,width=6.6in, height=9.1in]{geometry}
\usepackage{booktabs}

\newcommand{\calX}{\mathcal{X}}
\newcommand{\calA}{\mathcal{A}}
\newcommand{\calD}{\mathcal{D}}

\begin{document}

\title{Detection of False Data Injection Attacks (FDIA) on Power Dynamical Systems With a State Prediction Method\\
}

\author{\IEEEauthorblockN{Abhijeet Sahu}
\IEEEauthorblockA{\textit{Cyber security center} \\
\textit{NREL}}\\

\and
\IEEEauthorblockN{Truc Nguyen}
\IEEEauthorblockA{\textit{Computational Science Center} \\
\textit{NREL}}\\
\and
\IEEEauthorblockN{Kejun Chen}
\IEEEauthorblockA{\textit{Computational Science Center} \\
\textit{NREL}}\\

\and
\IEEEauthorblockN{Xiangyu Zhang}
\IEEEauthorblockA{\textit{Computational Science Center} \\
\textit{NREL}}\\

\and
\IEEEauthorblockN{Malik Hassanaly}
\IEEEauthorblockA{\textit{Computational Science Center} \\
\textit{NREL}}\\

}

\maketitle

\begin{abstract}
With the deeper penetration of inverter-based resources in power systems, false data injection attacks (FDIA) are a growing cyber-security concern. They have the potential to disrupt the system's stability like frequency stability, thereby leading to catastrophic failures. Therefore, an FDIA detection method would be valuable to protect power systems. FDIAs typically induce a discrepancy between the desired and the effective behavior of the power system dynamics. A suitable detection method can leverage power dynamics predictions to identify whether such a discrepancy was induced by an FDIA. This work investigates the efficacy of temporal and spatio-temporal state prediction models, such as Long Short-Term Memory (LSTM) and a combination of Graph Neural Networks (GNN) with LSTM, for predicting frequency dynamics in the absence of an FDIA but with noisy measurements, and thereby identify FDIA events. For demonstration purposes, the IEEE 39 New England Kron-reduced model simulated with a swing equation is considered. It is shown that the proposed state prediction models can be used as a building block for developing an effective FDIA detection method that can maintain high detection accuracy across various attack and deployment settings. It is also shown how the FDIA detection should be deployed to limit its exposure to detection inaccuracies and mitigate its computational burden.
\end{abstract}


\section{Introduction
}
\label{sec:introduction}
Electric grid failures can often be linked to frequency deregulation, particularly in instances where there is a lack of coordination among different operators, or when cyber intrusions lead to an imbalance between supply and demand. 
For instance, during the \textit{Western Systems Coordinating Council (WSCC)} event on February 7, 1996, a combination of events, including generation tripping and an inadequate response to extreme weather events led to a major blackout affecting the western United States and parts of Canada~\cite{1996_blackout}. 
Similarly, the \textit{Northeast Blackout of 1965} was triggered by a relay misoperation in Ontario, Canada, which caused a cascading failure that eventually led to the collapse of the entire interconnected grid. While not a deliberate attack, it highlighted vulnerabilities in the power system's resilience, i.e. its ability to handle disturbances and recover from failures~\cite{osti_5244283}. Physical failures can be exacerbated by cyberattacks, including false data injection attacks (FDIAs). FDIAs are a particular type of attack that aims to cause disruptions in the operation of the power grid by affecting the feedback mechanism that controls the grid. This is typically carried out by modifying the measurements used by the mechanism that approximates the state of the grid 
\cite{stealth_attack}. These attacks are not only confined to the sensor measurements but can also manipulate the controller parameters and input signals \cite{nguyen2020electric}. A recent example of FDIA is the \textit{Ukraine Power Grid Failure}~\cite{ukraine_attack} where attackers used malware to remotely access and manipulate control systems~\cite{ukraine_attack_fdi}, causing substations to trip and disrupting power supply for hundreds of thousands of customers.


Given how critical FDIAs can be, it is essential to equip power systems with reliable FDIA detection solutions. Attack and defense solutions against a steady-state power systems model have been extensively studied in the past such as~\cite{10197314,7579185} in the context of steady-state estimation. 
To the authors' knowledge, similar investigations about the efficacy of state prediction methods to capture FDIAs have not been carried out in the context of dynamic power system models. 

The importance of modeling the dynamics of power system can be understood from two angles. First, to remain stealthy, an attacker could attempt to adopt a time-dependent attacking strategy. Understanding the physical effect and the stealthiness of this approach requires modeling the power system dynamics. Second, prior work has shown that a carefully crafted time-dependent FDIA strategy could also induce frequency dynamics instabilities of higher magnitude than steady FDIAs \cite{rlfdi_prev_paper}.

Power system dynamics deals with the transient and dynamic behavior of the equipment such as the stabilizers systems and the machine models of the generators running in the power system. There, changes or disturbances in the operating conditions are reflected through fluctuations in the state variables such as rotor speed and angle that are not captured in steady-state systems. In turn, the effect of an FDIA can be more complex in a dynamical system than in a steady-state system. Hence, the objective of this work is to propose a detection method for FDIA, targeting power dynamical systems.


Traditionally, FDIA detection methods use state-prediction methods to identify whether an unplanned disturbance took place, such as an FDIA \cite{state_predict_1}. The state predicted is the system's state to be expected under nominal operating conditions. If the actual observations deviate from the predicted state, an unplanned event such as an FDIA likely occurred. This approach is advantageous in that any anomaly can theoretically be detected, irrespective of its nature \cite{state_predict_hmm}. In practice, detecting an anomaly requires carefully calibrating the magnitude of state discrepancies that can be tolerated \cite{stealth_attack}. Furthermore, to ensure that the detection task is computationally efficient, a data-based approach \cite{state_predict_lstm} rather than a physics-based approach \cite{state_predict_kalman} may be adopted. A data-based approach is especially beneficial when complex state dynamics need to be predicted, as is the case here. However, in the presence of an unplanned disturbance, the input of the data-based state predictor can be expected to fall outside of its training data distribution, thereby posing the question of their applicability after all.

Existing FDIA detection methods only determine whether the current state is under attack, which implies that the detection must be deployed at every single timestep. Although such a deployment is appropriate in a steady-state estimation, where the interval between timesteps are in the order of a few seconds or minutes, it is not efficient in dynamical systems where the interval is smaller (on the order of $10^{-3}$s \cite{steady_vs_dynamic}). Therefore, an appropriate FDIA detection method must also be computationally efficient in the case of dynamical systems.

In this work, several data-based state-prediction approaches are used to detect FDIA detection. The detection accuracy of these approaches under several FDIA types is evaluated and discussed. The novel contributions  of this work are as follows: 

\begin{enumerate}
\item A long-short term memory (LSTM) and a combined Graph Neural Network (GNN)-LSTM state predictor are demonstrated for the power dynamical system.
\item The effect of observation noise, dataset size, network architectures, and training sample size used on the state prediction performances are evaluated.
\item The FDIA detection accuracy using the state-predictors is evaluated for different FDIA schedules.
\item Recommendations for the use of data-based state predictors in detecting FDIAs are formulated

\end{enumerate}

In the rest of the paper, prior work on state-prediction methods for FDIA detection is reviewed in Sec.~\ref{sec:background}. The specific problem targeted is described in Sec.~\ref{sec:problemstatement}. Section~\ref{sec:lstm_gnn} describes the state-prediction methods and discusses their accuracy. The state-prediction methods are deployed to detect FDIA injection attacks in Sec.~\ref{fdi_truc} and the efficacy of the approach is discussed in Sec.~\ref{sec:eff_disc}.

\section{Prior Works/ Background}
\label{sec:background}

State estimation tasks for dynamic power systems are traditionally performed using a Kalman filtering approach~\cite{kf_1,kf_2}.
In Ref.~\cite{kf_1}, an iterated extended Kalman Filter (KF) based on the generalized maximum likelihood approach for estimating power system state dynamics is proposed, where the system considered can be subject to disturbances such as voltage collapse or line faults. A key deficiency noted is that the approach may be inappropriate under strong non-linearities of the system dynamics. Non-linearities can be better handled with Unscented Kalman Filters (UKF) \cite{maybeck1990kalman}. However, UKFs are notoriously sensitive to the number of sigma points needed to perform the unscented transform~\cite{ukf_deficit}, which can be problematic as the system dimensionality increases.  In \cite{kf_2}, a UKF approach that addresses numerical stability issues of UKF is proposed and demonstrated on a WSCC 3 machine and an NPCC 48 machine model. However, the approach is noise sensitive i.e. if the noise covariances are poorly estimated or if the noise change over time, the UKF is found to be unstable.  

Despite significant advances in the design of KFs, a key limitation remains about their computational cost for high-dimensional problems~\cite{kalman_comp_exp}, and their robustness to non-Gaussian noise \cite{non_gaussian_limitations}. Another challenge with the use of KF is the need to for accurate approximation of the system dynamics~\cite{kf_sys_approximation}.
By comparison, data-based approaches are more scalable with respect to the problem size and do not involve as many assumptions and system approximations~\cite{dd_no_approx}.

Unlike the Kalman filtering approach for state prediction, the model-free data-driven approach relies on historical data to make predictions. For instance, a data-driven power system state estimation is presented in~\cite{dd_state_predict} and such approaches are not only considered for future state predictions but for identifying anomalies in the pattern and intrusion detection. For instance, an attention-based auto-encoder approach to detect stealthy FDIA against the power system state estimation in an IEEE 14 system is proposed in~\cite{a3d}. Ref~\cite{power_dist_lstm} proposed a modified LSTM implementation for state estimation on a hybrid AC/DC distribution system composed of the IEEE 34-bus AC test system and a 9-bus DC microgrid.
Similarly, a probabilistic LSTM-autoencoder for solar power forecasting for intra-day electricity market was proposed~\cite{lstm_forecast_market}.

The state predictor can then be used for FDIA detection. For instance, Ref.~\cite{gnn} proposes using a GNN to detect stealthy FDIA against the power system state estimation in 3 different power transmission systems (IEEE 14, 118 and 300 systems). The stealthy attack implements a stochastic gradient descent (SGD) approach of tampering with the state variables, $V$ and $\theta$, of the neighboring bus from the point of attack bus using a breadth-first search approach, so that the measurement deviation remains under a particular threshold.
Unlike the present work, however, the temporal aspect of dynamical system is not effectively considered. Similarly, ~\cite{gnn_fdia,gnn_fdia_2,gnn_fdia_3} proposed different flavors of GNN for detecting intrusions and FDIA.
Though there are advancements in data-driven approaches for state prediction and FDIA detection, these approaches are only evaluated for steady-state systems.

\begin{figure}
    \centering
    \includegraphics[width=1.0\linewidth]{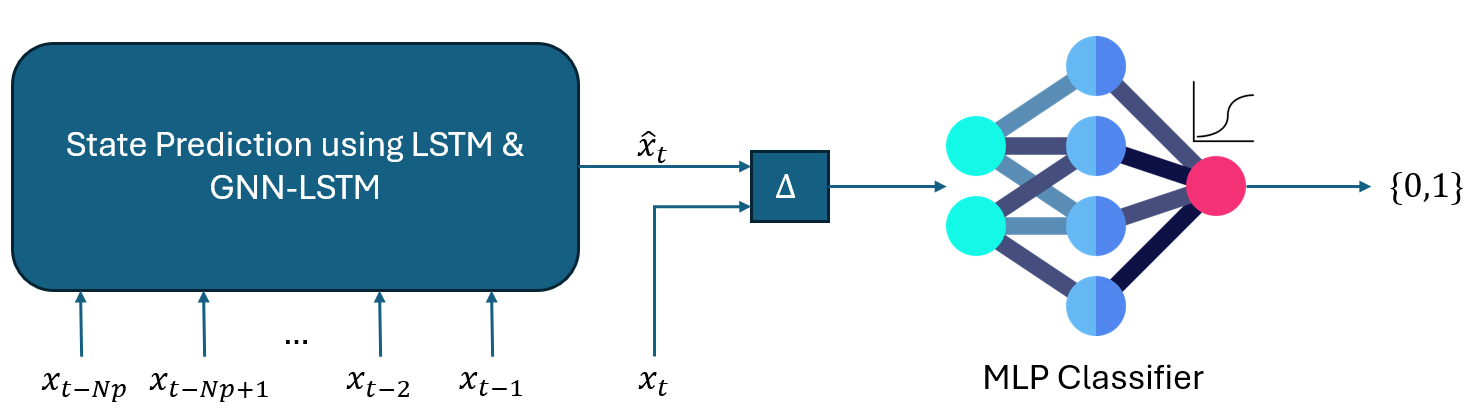}
    \caption{State-prediction based FDIA detector}
    \label{overall_architecture}
    \end{figure}

In this work, we propose the use of data-driven-based state-prediction for FDIA detection using deep neural networks for power dynamical systems. The architecture proposed in this work constitutes of a state predictor followed by a classifier, as shown in Fig.~\ref{overall_architecture}. We propose the use of LSTM and GNN-LSTM state predictor (Section~\ref{sec:lstm_gnn}) and use the prediction errors of these predictors as inputs to a classifier introduced in Section~\ref{fdi_truc}. 

\section{Problem statement 
}
\label{sec:problemstatement}
The problem of interest is in ensuring the stability of frequency dynamics via a primary frequency controller. Frequency dynamics is an important concern when integrating distributed energy resources (DER) which are often inverter-based and are characterized by a lower inertia \cite{denholm2020inertia}. The physical problem is modeled using the swing equation~\cite{kundur2022power}, here formulated as:
\begin{subequations}  \label{eq-swing-equation}
\begin{align}
    \dot{\theta_i} &= \omega_i,\\
    M_i\dot{\omega_i} &= p_{i}-p_{e, i}-D_i \omega_i-p^{\text{G}}_i
\end{align}
\end{subequations}
where, $i\in\mathcal{N}=\{1,..., N_b\}$ is the bus index, $\theta_i$ and $\omega_i$ are the voltage phase angle and frequency deviation of bus $i$ respectively, and are the state variables of the swing equation. $M_i$ and $D_i$ are the inertia and damping coefficients. The net power injection at bus $i$ is denoted by $p_i$. $p^{\text{G}}_i$ represents the power output from the generator on bus $i$.
To maintain the system’s frequency stability, the fast-responding Inverter-based Resources (IBR) on all $N_b$ buses participate in the system's primary
frequency control following their designed droop rule,
i.e., $p_i^{G}= k_i\;\omega_i$, where $k_i$ is the droop controller coefficient of the generator at bus $i$.

In this work, the goal is to first predict the future state of the state variables $\theta_i^{t+1}$ and $\omega_i^{t+1}$, given an observation window length, $N_p$ of prior observations $\theta_i^{t-N_p,...,t}$ and $\omega_i^{t-N_p,...,t}$. In practice, $N_p>0$ because of possible noise in the data acquired by phasor measurement units (PMU) \cite{brown2016characterizing}. Then by using the error between the actual measurements and the predicted state values, one would detect whether an FDIA occurred. The hypothesis behind the state prediction approach for FDIA detection is: \textit{State predicted based on an un-perturbed or an un-attacked system will deviate from the actual state on an attacked system}. The FDIA considered in this work is one that affects the droop control coefficient $k_i$ to perturb the system dynamics. The system considered throughout this work is the Kron-reduced IEEE New England Transmission System with $N_b=10$ buses on which both traditional synchronous machines and IBRs exist. The physics model is implemented using the available implementation of Ref.~\cite{cui2022reinforcement}. 

In Sec.~\ref{sec:lstm_gnn}, the state prediction approaches adopted and evaluated are further described. The trained state-prediction models paired with a classifier are then leveraged to detect FDIAs. The type of FDIAs considered, as well as the construction of the classifier are described in Sec.~\ref{fdi_truc}. The overall pseudo code for the approach adopted for prediction followed by detection is shown in Algorithm~\ref{alg:detection}.

\section{State prediction for power system dynamics}
\label{sec:lstm_gnn}

In this section, the objective is to provide guidance in the design of state prediction approaches for power system dynamics. Two state prediction approaches are discussed and evaluated with different network architectures and under varying noise levels.


\subsection{LSTM}
Long Short-Term Memory (LSTM) networks are a type of recurrent neural network (RNN) architecture designed to address the vanishing gradient problem in traditional RNNs \cite{hochreiter1997long}. The gated architecture of LSTMs (comprising the input, forget, and output gate) regulates the flow of information within the network, enabling it to selectively remember or forget information from previous timesteps based on the current input. In turn, this helps LSTMs retain information over longer time intervals than RNNs. In the present case, a long-term memory is necessary to combat possible noise in the observed system state variables.

Here, an LSTM autoencoder \cite{srivastava2015unsupervised} is used instead of plain LSTM. LSTM autoencoders can effectively learn compact representations of input sequences by compressing them into a lower-dimensional space that captures the most important features. LSTM autoencoders were observed to lead to better performance in tasks such as sequence generation~\cite{lstm_ae_seq_gen}, anomaly detection~\cite{nguyen2021forecasting}, and denoising~\cite{lstm_ae_denoising}, which are all essential here. 

\begin{figure}
    \centering
    \includegraphics[width=1.0\linewidth]{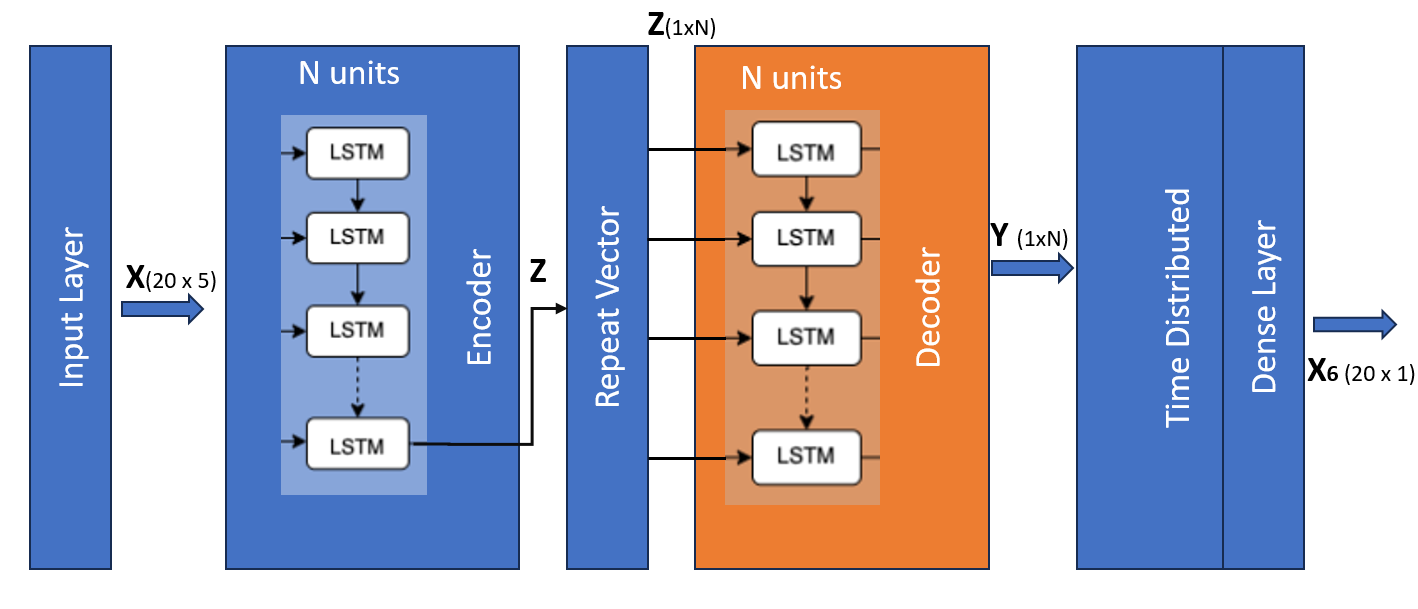}
    \caption{LSTM architecture with $N_f$ = 20 and $N_p$ = 5}
    \label{lstm_architecture}
    \vspace{-5mm}
\end{figure}

The present implementation consists of encoder and decoder layers. The input dimension to the encoder is $Z \in \mathcal{R}^{N_{p} \times N_{f}}$, where $N_{p}$ is the number of past sequence and $N_{f}$ is the number of features. The features are the state variables $\theta$ and $\omega$ of the ten generator buses in the IEEE 39 New England Kron-reduced model. The number of LSTM units for the encoder layer is defined through $L_{u}$. The output from the encoder is fed into a decoder LSTM layer with the same number of  LSTM units, $L_{u}$. The output is further passed through a dense layer of units $N_f$ and a \textit{TimeDistributed} layer, which is a wrapper to apply a layer to every temporal slice of an input. Further, the model is compiled through the Adam optimizer \cite{kingma2014adam} and a mean square error loss function is used. The dataset uses a 70-30 training/validation split and the model is trained for $N_{e}$ epochs. A \textit{tanh} activation and a \textit{sigmoid} recurrent activation are used in the respective encoder and decoder LSTM layers. The LSTM auto-encoder architecture is shown in Fig.~\ref{lstm_architecture}. Here, $N_f=20$ since there are 10 buses in the reduced model and each bus has two state variables $\omega$ and $\theta$. Throughout all the experiments, $N_p=5$.

\subsection{GNN-LSTM}
Unlike LSTMs, Graph Neural Networks (GNN) are neural networks designed to work with data structured as a graph \cite{battaglia2018relational}. GNNs leverage the graph structure of the problem to encode the necessary inductive bias. They allow performing tasks such as node classification \cite{gnn_node_class}, link prediction \cite{gnn_link_predict}, graph classification \cite{gnn_graph_class}, and graph generation \cite{gnn_graph_gen}. Given the nature of the present problem, it is natural to consider combining the benefits of a GNN with the benefits of an LSTM for state prediction. The implementation adopted here is based on~\cite{gnn_lstm_predict}, originally applied to traffic forecasting. 

The physical system considered here is modeled with a connected, undirected, weighted graph $\mathcal{G}=(\mathcal{V},\mathcal{E},\textbf{W})$ that consists of a finite set of vertices $\mathcal{V}$ with $|\mathcal{V}| = N_b$, where $N_b$ is the number of buses, a finite set of edges $\mathcal{E}$ and a weighted adjacency matrix $\textbf{W} \in \mathcal{R}^{N_b \times N_b}$. If the buses $i$ and $j$ are connected, the corresponding weight of the edge $e = (i,j)$ connecting vertices $i$ and $j$ is assigned to $W_{ij}$. A signal or a function $f : \mathcal{V} \rightarrow \mathcal{R} $ can be represented by a vector $\textbf{\textit{f}} \in \mathcal{R}^{N_b}$, where $i^{th}$ component of the vector \textbf{\textit{f}} corresponds to state value at the vertex/bus $i \in \mathcal{V}$. 
The adjacency matrix considered for the $\mathcal{G}$ is based on the Kron-reduced IEEE 39 model considered in~\cite{ref_kron}.

The input data is passed through a Graph Convolution Network (GCN) Layer where a message passing operation is performed. GNNs typically operate through message-passing schemes where information is propagated between neighboring nodes in the graph. At each step of message passing, nodes aggregate information from their neighbors and update their own representations based on the aggregated information (Line 10 of Alg.~\ref{alg:detection}).  

There are three common types of aggregation considered: a) sum, b) mean, and c) max. In \textit{mean aggregation}, the representations of neighboring nodes are averaged to compute the updated representation of the central node. It is simple and computationally efficient~\cite{gnn_agg_mean}. However, it may discard important information, especially in graphs where some neighboring nodes carry more relevant information than others. The \textit{sum aggregation} sums the representations of neighboring nodes. It is computationally efficient and retains all the information from the neighborhood. However, it suffers from the problem of oversmoothing~\cite{gnn_agg_sum}, where the representations of nodes become too similar after multiple message-passing steps, leading to loss of discriminative power. The \textit{max aggregation} selects the maximum value from the representations of neighboring nodes. Such aggregation can capture the most relevant information from the neighborhood, making it robust to noise. However, it does not appropriately handle cases where multiple neighboring nodes carry relevant information~\cite{gnn_agg_max}. These three aggregation types are evaluated in Sec.~\ref{aggregation_eval}. Finally, in the update stage, the node representation are updated by either concatenation or addition operation. 

\begin{figure}
    \centering
    \includegraphics[width=1.0\linewidth]{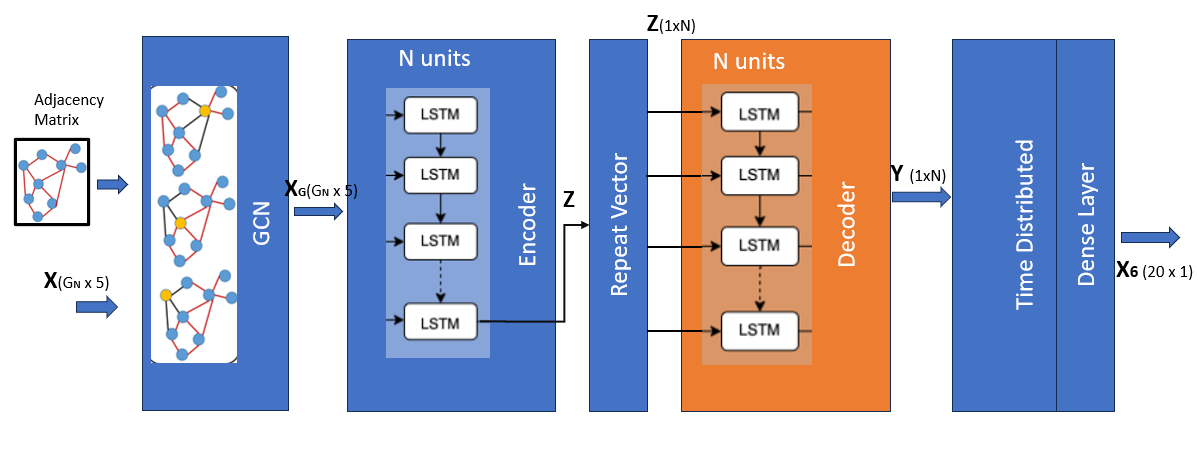}
    \caption{GNN-LSTM architecture where we feed the adjacency matrix of the power topology alongwith the observation of $N_p$ = 5.}
    \label{gcn_lstm_architecture}
\end{figure}

To process the time-dependent information, the outputs of the GCN is passed into LSTM encoder and decoder layers. The training procedure uses the same optimizer, loss, train/test splits and activation in the LSTM layers. The combined GCN followed by the auto-encoder-based LSTM architecture is illustrated in Fig.~\ref{gcn_lstm_architecture}.

\begin{algorithm}[t]
\begin{small}
  \caption{Pseudo Code: State prediction-based False Data Injection Attack Detection}\label{alg:detection}
	\begin{algorithmic}[1]
	\State{Define $P_{type}$, $L_u$, $N_p$, $N_f$, $Agg$}
	\For{$s$ = [500,1000,1500]} \Comment{Vary sample size for training}
	    \For{$\sigma^{2}$ = [0,0.001,0.005,0.01]} \Comment{Vary Noise $\mathbb{N}(0,\sigma)$}
            \State Add Gaussian Noise to measurement
            \State Split dataset to training and testing
            \If { $P_{type}==LSTM$}
                \State Pass through the model (Fig.~\ref{lstm_architecture})
            \Else \Comment{Consider the combined GNN-LSTM architecture}
                \State Compute the weighted  adjacency matrix $W$
                \State Perform aggregation ($Agg$) and combination operation within the GCN layer
                \State Pass through the model (Fig.~\ref{gcn_lstm_architecture})
            
            \EndIf
            \State Compile and fit the state-predictor model $g_\theta$
	      \State Predict the state and compute the prediction error.
            \State Perform FDIA attacks: Sliding window and Cyclic \Comment{Refer to Section.~\ref{detection_result}}
            \State Consider prediction error to train FDIA detection classifier, $h_\phi$.
            \State Evaluate based on Accuracy, F1-score, Precision, Recall of the $h_\phi$ under varying attack types and noise levels.
	    \EndFor
	\EndFor
  \end{algorithmic}
  \end{small}
\end{algorithm}

\subsection{Dataset}
\label{sec:dataset}
The dataset considered comprises the state variables, phase angle ($\theta$), and frequency ($\omega$) of the $N_b =10$ buses in the IEEE 39 New England Kron-reduced model. This dataset is created using solution of the swing equation (Eq.~\ref{eq-swing-equation}). The configuration adopted corresponds to the environment used in a reinforcement learning (RL) environment developed by the same authors \cite{rlfdi_prev_paper} and is available upon request.
The dataset is assembled by simulating so-called ``episodes" which echoes the RL implementation mentioned above. Each episode in the environment comprises of 500 timesteps of size 0.01s each. The rationale behind the use of episodes is that the state-predictor should be equally capable during the overshoot phase and the stable phase of the system. Both phases are captured over the time interval (5s) considered here. In total, 10,000 episodes are generated by sampling the initial values of $\omega$ and  $\theta$ from the uniform distributions of $\mathbb{U}(0,0.3)$ and $\mathbb{U}(-0.03,0.03)$ respectively. The states are predicted for every $(N_p + 1)^{th}$ time instance for every $N_p$ past temporal state values. The observations are superimposed with varying levels of noise as typically observed in the Phasor Measurement Units (PMU) \cite{brown2016characterizing}. The noise is assumed to be additive and normally distributed as $\mathbb{N}(0,\sigma)$ where $\sigma$ is the standard deviation of the noise, also referred to as noise level. The noise level $\sigma$ is varied in the set $\{0, 0.001, 0.005\}$ in the rest of the paper. The same noise levels (defined by $\sigma$) are used for each bus and state variable.

\subsection{Result}
Hereafter, the performances of the predictor are evaluated on the basis of mean absolute error (MAE) and mean relative error (MRE). The MAE and MRE are computed using the following equations:
\begin{equation}
    MAE(X) = \frac{\sum_{i=1}^{D} \sum_{j=1}^{N_b}|x_{i,j,\textrm{act}}-x_{i,j,\textrm{pred}}|}{D \times N_b}
\end{equation}
\begin{equation}
    MRE(X) = \frac{\sum_{i=1}^{D}\sum_{j=1}^{N_b}\frac{|x_{i,j,\textrm{act}}-x_{i,j,\textrm{pred}}|}{x_{i,j,\textrm{act}}}}{D \times N_b}
\end{equation}
where, $D$ is the number of data points in the testing set, $x_{i,j,\textrm{act}}$ and $x_{i,j,\textrm{pred}}$ refers to the actual and predicted state variables for the $i^{th}$ datapoint and the $j^{th}$ bus. In the following, $x$ can refer to either frequency or phase which is clarified by the use of the notations MAE($\theta$) and MAE($\omega$). The MAE is considered since it is robust to outliers as compared to MSE. Note that the MRE can be useful to evaluate the relative accuracy but can also be misleading when the actual values are nearing zero. MRE is still reported for completeness. 

\subsubsection{Evaluation of LSTM based prediction with non-noisy data and varying LSTM units}

In this section, no noise is applied to the dataset, and this section aims to understand what architecture choices are needed to capture the system dynamics. The first network architecture variable investigated is the number of LSTM units. Increasing the number of LSTM units increases the model's capacity at capturing complex dynamics. With more units, the network can potentially learn richer representations, which leads to improved accuracy, especially for tasks that require modeling intricate temporal dependencies.
For larger and more complex datasets, models with a greater number of units may be better able to capture the underlying patterns. However, for smaller datasets, increasing model capacity may also lead to overfitting.  Adding more LSTM units increases the computational complexity of the model, making training and inference more computationally demanding. In general, larger models require more training time, memory, and computational resources.

\begin{figure*}[h]
  \centering
  \subfigure[]{\includegraphics[width=0.46\linewidth]{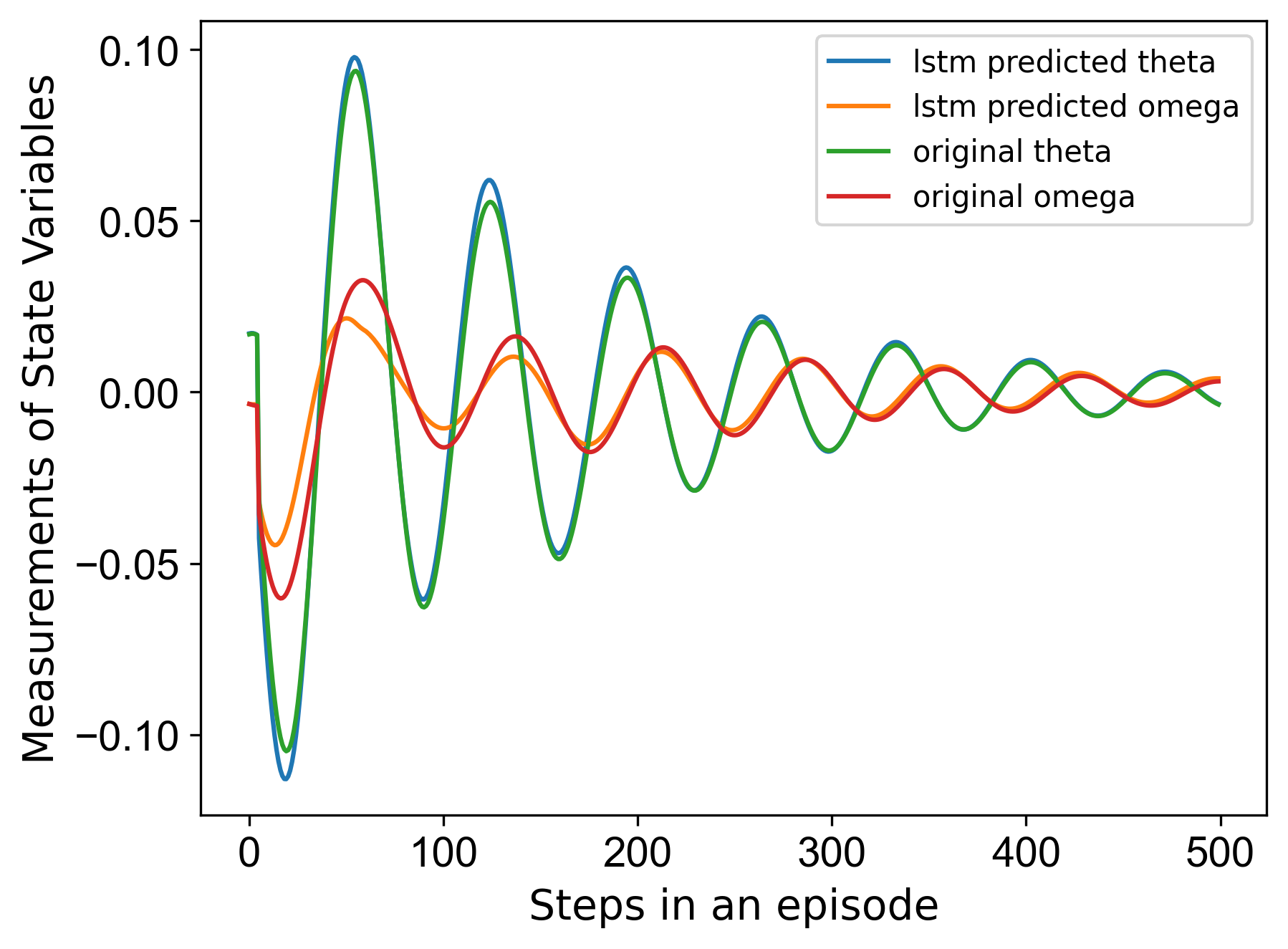}\label{pred_10}}\hspace{5mm}
  \subfigure[]{\includegraphics[width=0.46\linewidth]{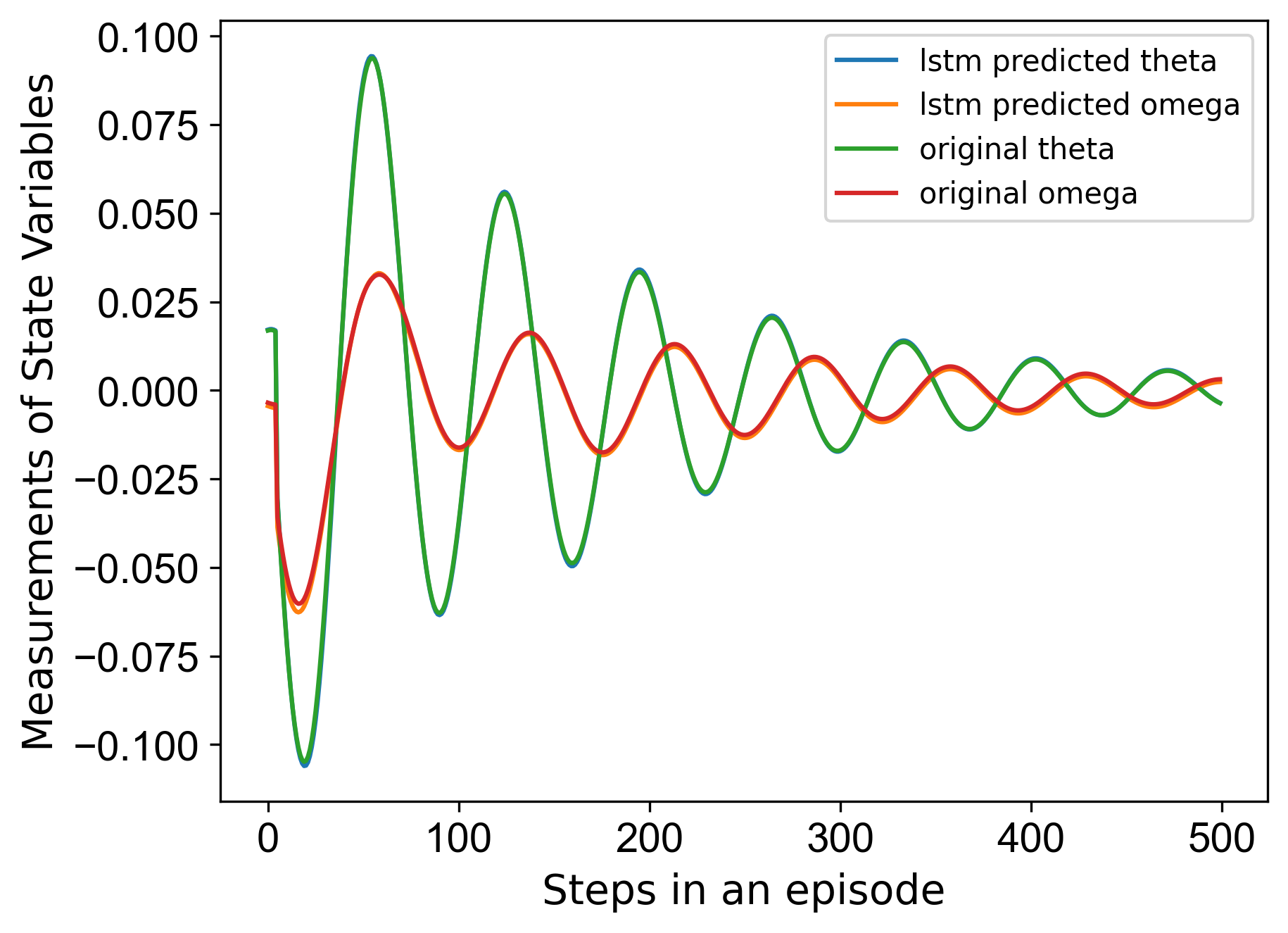}\label{pred_25}}
  \caption{LSTM prediction with varying LSTM units: a) Actual and predicted state with a) $L_u$=10 b) $L_u$=25}
\end{figure*}

\begin{figure}
    \centering
    \includegraphics[width=0.9\linewidth]{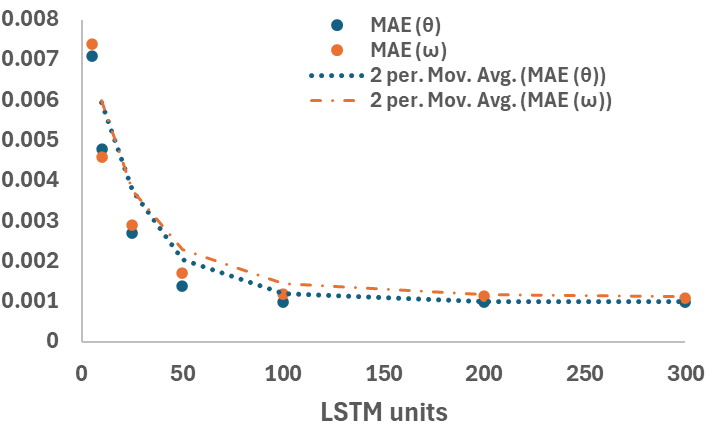}
    \caption{MAE for prediction with varying LSTM units}
    \label{fig:lstm_varying_units}
\end{figure}

Figure~\ref{pred_10} shows the predicted states
trained with LSTM model with $L_u=10$ units, while Fig.~\ref{pred_25} shows the predicted states when using with $L_u=25$ units. The x-axis in the plot refers to the number of steps in one episode of the swing equation environment (see Sec.~\ref{sec:dataset} for the terminology used here), while the Y axis refer to the actual and predicted states at Bus 1. Similar results are observed for other buses and are not shown for the sake of clarity. With $L_u=10$, higher errors are observed during the overshoot phase of the transient as compared to the case with $L_u=25$. Overall, improvements due to a higher number of units are mostly observed in the overshoot phase. Fig~\ref{fig:lstm_varying_units} shows the MAE of the state predicted averaged over all the 10 buses when the training is performed with varying LSTM units trained with 25 epochs.  Improved accuracy is observed as the number of units increased. But the accuracy saturates further as the model size increases.

\subsubsection{Evaluation of GNN-LSTM with different aggregation approach}\label{aggregation_eval}

The aggregation Types considered for the GCN layer of the GNN-LSTM architectures are: a) Mean, b) Sum and c) Max.
Table~\ref{varying_aggregation} shows the MAE, MSE and validation loss obtained after training the GNN-LSTM state predictor through 25 epochs, using three different types of aggregation considered in the GCN layer. For the LSTM auto-encoder block, 25 LSTM units are considered for both encoder and decoder layers.
The \textit{max aggregation} performed the best among all the three types of aggregation. Notably, the fastest training convergence was obtained in the case of \textit{mean aggregation} (converged in 5 epochs) but it converged to a higher loss in comparison to the other aggregation approaches.

 \begin{table}[]
 \caption{Evaluation of prediction accuracy with varying Aggregation operation in the GCN layer}
\label{varying_aggregation}
\centering
\begin{tabular}{c c c c} 
 \hline
Aggregation Type & MAE & MSE & Validation Loss \\ [0.5ex] 
 \hline
Mean & 0.06 & 0.021 & 0.02\\ 
Sum & 0.049  & 0.011 & 0.01\\
Max & 0.048 & 0.012 & 0.01\\
 [1ex] 
 \hline
 \end{tabular}
\end{table}


\subsubsection{Evaluation of impact of noise} \label{lstm_noise}
In this section, the focus is on evaluating the state predictor's performance under noisy observations. The networks evaluated under different levels of noise are also retrained with data that has the same level of noise. The expected outcome is that the state-predictor model learns to ignore the noise and captures only the underlying dynamics.
Table~\ref{lstm_varying_noise} shows the result for MAE and MRE averaged over all the 10 buses for each state variable $\theta$ and $\omega$. The MAE error closely follows the noise level, suggesting that the LSTM predicts the conditional average of the dynamics, averaging out the noise. Once again, large MRE values for $\omega$ are observed because the frequency deviation is intermittently nullified.

\begin{table}[]
 \caption{Average MAE and MRE results for prediction using LSTM with varying noise levels}
\label{lstm_varying_noise}
\centering
\begin{tabular}{c c c c c} 
 \hline
Noise & MAE($\theta$) & MRE($\theta$) & MAE($\omega$) & MRE($\omega$) \\ [0.5ex] 
 \hline
$\mathbb{N}(0,0)$ & 0.000039 & 0.0025 & 0.000082 & 1.91\\ 
$\mathbb{N}(0,0.001)$ & 0.000894 & 0.0512 & 0.000875 & 3.129\\
$\mathbb{N}(0,0.005)$ & 0.0043 & 0.474	& 0.0043 &	3.66 \\
 [1ex] 
 \hline
 \end{tabular}
\end{table}


Table~\ref{gnn_varying_noise} shows the same results for the GNN-LSTM that uses the same number of units as the LSTM. While in the noiseless case, the MAE is improved as compared to the LSTM, the performances of the GNN-LSTM are also strongly affected by the noise level. The MAE is significantly degraded compared to the LSTM, for $\omega$ in particular.

\begin{table}[]
 \caption{Average MAE and MRE results for prediction using GNN-LSTM with varying noise levels}
\label{gnn_varying_noise}
\centering
\begin{tabular}{c c c c c} 
 \hline 
Noise & MAE($\theta$) & MRE($\theta$) & MAE($\omega$) & MRE($\omega$) \\ [0.5ex] 
 \hline 
$\mathbb{N}(0,0)$ & 0.0000295 & 0.0023 & 0.000075 & 1.45\\ 
$\mathbb{N}(0,0.001)$ & 0.00203 & 0.25 & 0.048 & 2.61\\
$\mathbb{N}(0,0.005)$ & 0.00815 & 0.88	& 0.196 &	4.62 \\
 \hline
 \end{tabular}
\end{table}

\begin{figure*}[h]
  \centering
  \subfigure[]{\includegraphics[width=0.46\linewidth]{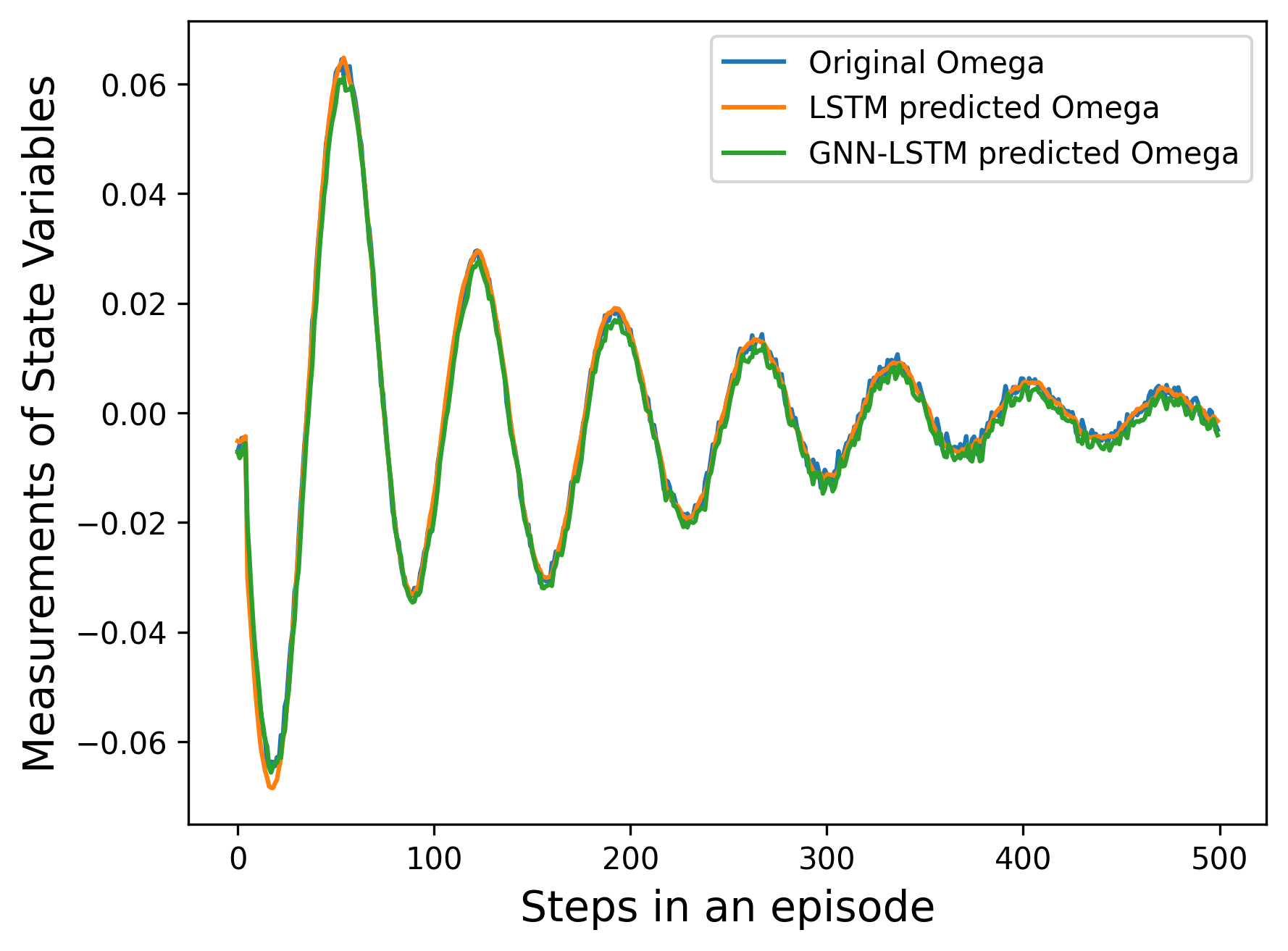}\label{pred_w_noise}}\hspace{1mm}
  \subfigure[]{\includegraphics[width=0.46\linewidth]{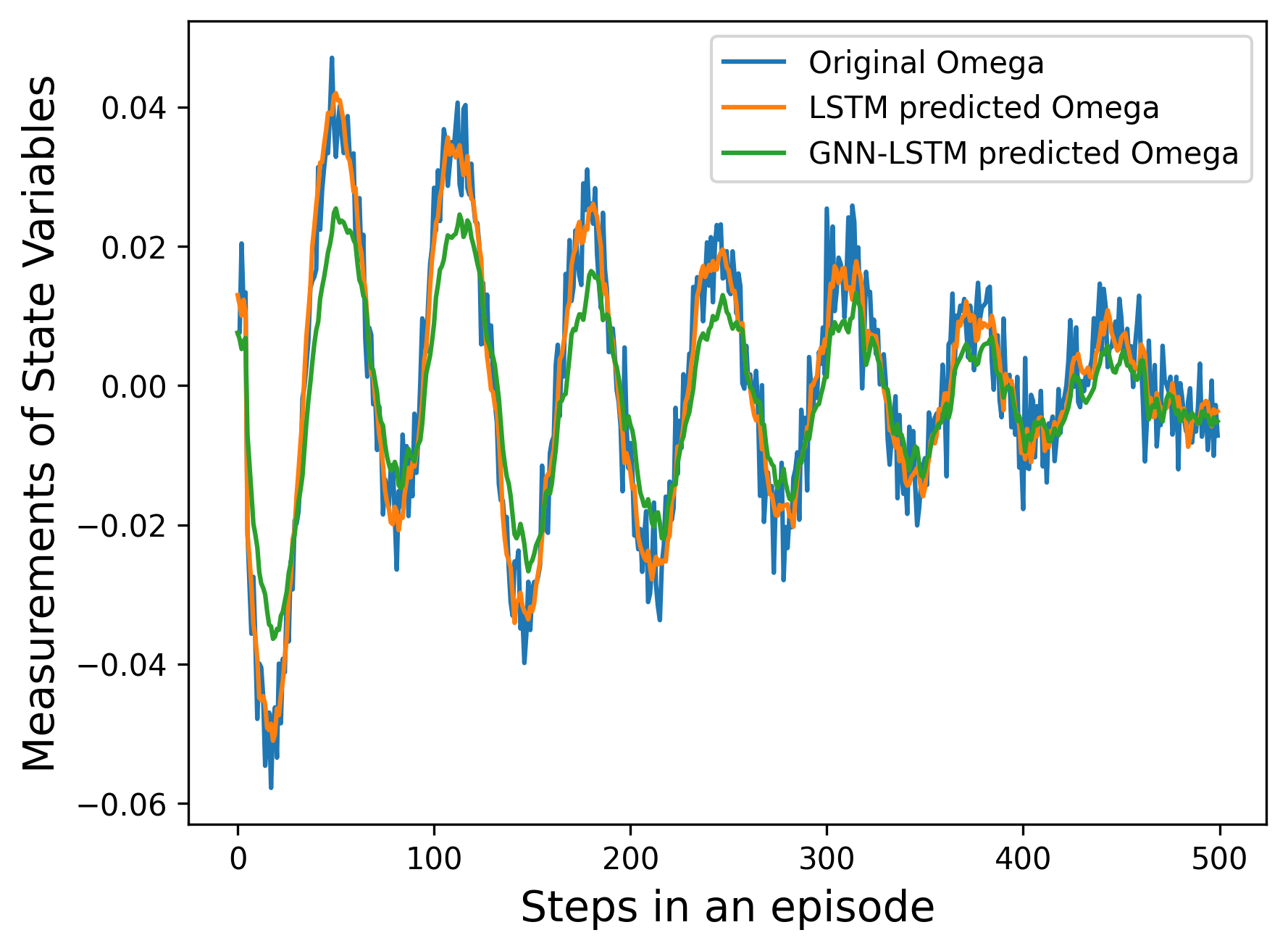}\label{pred_w_med_noise}}\hspace{1mm}
  \caption{Prediction comparison of $\omega$ state variable at Bus 0 of the IEEE 39 New England Kron-reduced model, with varying Gaussian Noise $\mathbb{N}(0,\sigma)$: with a) $\sigma$ = 0.001, b) $\sigma$ = 0.005}
\label{pred_noise}
\end{figure*}


The dynamics predicted by both the LSTM and the GNN-LSTM are shown in Figs~\ref{pred_noise}. It can be seen that while the LSTM manages to average out the noise, the GNN-LSTM also introduces a consistent bias in the dynamics. The bias is especially apparent for larger levels of noise (Fig.~\ref{pred_w_med_noise}), and during the overshoot phase.



\subsubsection{Evaluation of impact of training sample size}

The GNN-LSTM can be expected to be sample efficient compared to the LSTM given that it encodes the relational structure in the data. Similar findings have been observed in physics-informed approaches \cite{hassanaly2024pinn} for the same reason. In this section, the objective is to characterize the effect of the training data size for the LSTM and the GNN-LSTM, with and without noise. The number of units chosen is 25 for both. Out of the 10,000 episodes in the training dataset (Sec.~\ref{sec:dataset}), a subset is used and referred to as \textit{Sample}. The Sample values considered are $\{500, 1000, 1500\}$.
Table~\ref{varying_samples_table} show that in the absence of noise, the accuracy of the GNN-LSTM improves with sample size. In contrast, the accuracy of the LSTM fails to achieve reasonable accuracy levels and does require more training data. In the presence of noise ($\sigma=0.001$), the accuracy of the GNN-LSTM only slowly improves with sample size and the advantage of the GNN-LSTM over the LSTM is less clear. This result echoes the finding of Sec.~\ref{lstm_noise} where the inductive bias of the GNN-LSTM did not help in the presence of noise.



\begin{table}[]
 \caption{Average MAE with varying training sample-size used with non-noisy and noisy measurements.}
\label{varying_samples_table}
\centering
\begin{tabular}{c c c c c} 
 \hline 
Sample & LSTM & LSTM ($\mathbb{N}$) & GNN-LSTM & GNN-LSTM ($\mathbb{N}$)\\ [0.5ex] 
 \hline 
500 & 0.029 & 0.032  & 0.017  & 0.042\\ 
1000 & 0.032 &  0.035 & 0.014  & 0.038\\
1500 & 0.036 & 0.038	& 0.009 & 0.037\\
 \hline
 \end{tabular}
\end{table}

\subsection{Discussion}
From the experimental results, it can be inferred that the accuracy in the overshoot phase of the transient is affected by the number of LSTM units used in the encoder and decoder layers. It was observed that the GNN-LSTM predictor only improves the accuracy, compared to an LSTM, when trained with non-noisy measurements. This effect can be understood by the fact that the physical dynamics are occulted by the noise. In particular, the noise destroys the spatial correlation and the graph topology encoded in the GNN becomes ineffectual. Another hypothesis is that the aggregation method is inappropriate to successfully average out the noise. The design of a better aggregation method is left for future work. Additionally, with small data, the prediction accuracy of the GNN-LSTM increases faster than the plain LSTM. Therefore, for practical applications that can only use small datasets, the advantage of the GNN-LSTM could be significantly clearer. In the upcoming section, a classifier is trained for detecting FDIAs. It leverages both the LSTM and GNN-LSTM state predictors.

\section{FDIA Detection} 
\label{fdi_truc}
Leveraging the state predictors proposed in Sec.~\ref{sec:lstm_gnn}, a detection model is devised to detect FDIAs via inspecting prediction errors. 
This section begins with establishing a threat model that specifies the FDIA detection problem, describes the proposed detection model, and finally presents a suite of experiments demonstrating the detection's efficacy under various settings. 

\subsection{Threat Model}

\noindent\textbf{Adversary.} The FDI attacker is considered to be an intruder who has access to the control center of every inverter-based resource and can tamper with the droop coefficient of the controller. Denoting $\calA$ as an adversary who can perturb the droop coefficients $k_i^t$ of some bus $i$ at time $t$, the adversary's goal is to induce frequency oscillations while remaining undetected by the system.
Previous work \cite{rlfdi_prev_paper} has shown that, by setting $k_i^t=-1$ of Eq.~\ref{eq-swing-equation}, the attack will result in damaging frequency oscillations. The adversary $\calA$ is represented as a function $\calA(t) \rightarrow \{0, 1\}$ that decides whether to perturb the droop coefficient at timestep $t$. If $\calA(t) = 1$, the adversary sets the droop coefficient to $-1$.

\noindent\textbf{Detection.} 
The goal of detection is to determine whether an FDIA has occurred. Note that the system does not have direct access to the values of $k_i^t$ but only to the measurements (i.e., frequencies and phases) at each timestep. For simplicity, the measurements of all buses are flattened at timestep $t$ into a one-dimensional vector $x\in \mathbb{R}^d$.
The detection is represented as a function $\calD(x_{t-N_p}, x_{t-N_p+1},...,x_{t}) \rightarrow \{0,1\}$, where $x_t$ is the state observed as time $t$. 
A successful detection method returns $1$ if an attack occurs \textit{at any} of the timesteps from $t-N_p$ to $t$, and $0$ otherwise. 
%

\begin{figure}[h]
    \centering
    \includegraphics[width=0.9\linewidth]{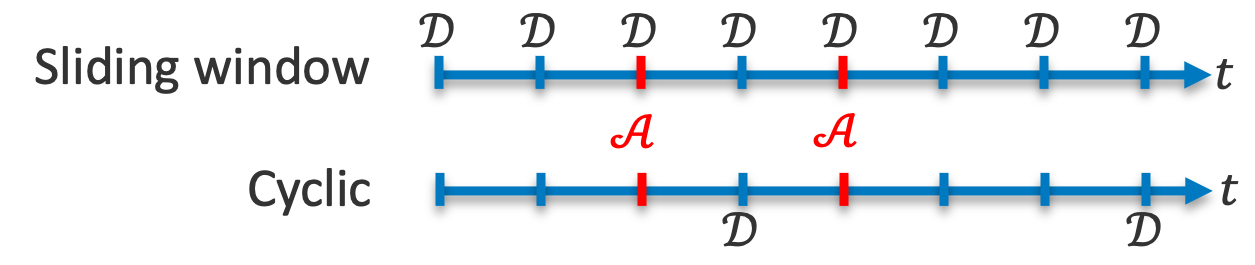}
    \caption{Two deployment settings for $\calD$. In sliding window, the inference step would be perturbed in case of an FDIA. In cylic, the perturbation of an FDIA might happen in the observation window, not the inference.}
    \label{fig:deployment}
\end{figure}

This formulation of $\calD$ enables two deployment settings for the detection method: \textit{sliding window} and \textit{cyclic}, as illustrated in Fig. \ref{fig:deployment}. In the \textbf{sliding window} setting, the detection method is deployed at every single timestep, signifying that if an FDIA happens, the inference step $t$ will be perturbed. This setting is often considered in previous work on bad data detection and FDIA detection in power systems \cite{10317877}, where the goal is only to determine whether the current timestep $t$ is under attack or not. In the \textbf{cyclic} setting, the detection is deployed at some fixed interval such that an FDIA might not always result in perturbing the inference step $t$. This means the detection must also be able to detect the case that an attack does not occur at the inference step $t$ but in the observation time window $\{t-N_p,...,t-1\}$. Since the latency between consecutive timesteps is short in dynamic-state systems (as discussed in Section \ref{sec:introduction}), a cyclic deployment is more cost-effective and practical than a sliding window deployment. 

\subsection{Detection Method} \label{ssec:detect}
The core challenge in realizing the detection function $\calD$ is that it needs to detect attacks that happen over a time window rather than at the current timestep. 
To achieve this, the proposed state prediction (Sec.~\ref{sec:lstm_gnn}) is leveraged as a building block to develop an FDIA detection model as follows.

The state predictor (e.g., LSTM or GNN-LSTM) is abstracted as a function $g_\theta: \calX^{N_p} \rightarrow \calX$ parameterized by a set of weights $\theta$ that predicts $\hat{x}_t = g_\theta(x_{t-1}, x_{t-2},...,x_{t-N_p})$ where $\hat{x}_t, x_t \in \mathbb{R}^d$ denote the predicted and actual observations at time $t$, respectively, and $N_p$ denotes the size of a time window. After obtaining the actual observations at time $t$, the prediction error is then calculated as $e_t = (\hat{x}_t - x_t)^2 \in \mathbb{R}^d$. Next, a binary classifier $h_\phi: \mathbb{R}^d \rightarrow [0,1]$ is trained to map the prediction error $e_t$ to a score that indicates the probability of an attack happening during the observed time window. The score is then compared with a threshold of 0.5 to obtain the predicted label at time $t$. The detection method $\calD$, hence, is \textit{a composition of the state prediction model and the binary classifier}, i.e., $\calD = h_\phi \circ g_\theta$.

By leveraging the prediction error, such a detection function $\calD$ would determine whether the measurement at the inference timestep $t$ is abnormal with respect to the measurements of the observation time window $\{t-N_p,...,t-1\}$. Intuitively, assuming that the attack happens at $t$, the observed state $x_t$ would be deviating from the predicted state $\hat{x}_t$, causing high prediction error $e_t$. On the other hand, if an attack happens at any timestep in the observation window $\{t-N_p,...,t-1\}$, the state prediction model $g_\theta$ would be affected by such attacks, resulting in a "poisoned" prediction $\hat{x}_t$ that would also cause high prediction error. Therefore, such a detection method should be able to catch attacks at any timestep from $t-N_p$ to $t$.



\subsection{Results}
\label{detection_result}
\newcommand{\calW}{\mathcal{W}}

\subsubsection{Experimental settings}
A window-level detection with a window size of $N_p + 1$ ($N_p$ timesteps in the observation window, and 1 inference timestep) is considered. At the current timestep $t$, a time window $\calW \equiv \{t-N_p,...,t\}$ is attacked (or adversarial) if there exists an adversarial timestep in the window, i.e., $\exists j \in \calW: \calA(j) = 1$, and is benign otherwise. Given a time window $\calW$, the detection succeeds if it can correctly predict whether $\calW$ is adversarial or benign.

To evaluate the performance of the FDIA detection, the swing equation (Eq.~\ref{eq-swing-equation}) is integrated for the 10-bus IEEE 39 New England Kron-reduced model to generate a dataset containing benign and adversarial windows $\calW$. Each adversarial window is created by perturbing $m \in [1, N+1]$ random timesteps in the window. For simplification, the attack is systematically conducted on bus 7 which is the bus that induces damaging frequency oscillations on the system \cite{rlfdi_prev_paper}.  Section~\ref{sec:multibus} later extends the experiments to multiple buses. The dataset contains $200,000$ windows with $100,000$ benign and $100,000$ adversarial windows. The dataset is further divided into train/test splits at $80/20$ with each of them having an equal number of benign and adversarial windows. In the following experiments, the observation window size $N_p$ is set to 5, consistent with Sec.~\ref{sec:lstm_gnn}. With this dataset, the baseline accuracy of detection is 0.5, which is the probability of random guessing.


Multiple state prediction models $g_\theta$ are evaluated, including two LSTM models with 100 and 50 units, respectively, and one GNN-LSTM model with 25 units. These models are trained as described in Section \ref{sec:lstm_gnn}. Using several predictor models allows for characterizing the effect of the state predictor accuracy on the detection accuracy. On the other hand, $h_\phi$ is implemented using a multilayer perception architecture and is trained on the generated dataset using an Adam optimizer. With respect to each implementation of $g$, the hyperparameters of $h_\phi$ are tuned using Optuna \cite{akiba2019optuna} and presented in Table \ref{tab:param}. More details regarding the tuning process can be found in the Appendix.

\begin{table}[h]
\caption{Hyperparameters of $h_\phi$}
\label{tab:param}
\begin{tabular}{@{}llll@{}}
\toprule
\multicolumn{1}{c}{\multirow{2}{*}{\textbf{$g_\theta$}}} & \multicolumn{3}{c}{\textbf{Hyperparameters of $h_\phi$}}                \\ \cmidrule(l){2-4} 
                                     & \textbf{\# hidden layer} & \textbf{\# neurons} & \textbf{Learning rate} \\ \midrule
LSTM ($L_u = 50$)                    & 1                        & 20                  & 0.00376                \\
LSTM ($L_u = 100$)                   & 1                        & 30                  & 0.0034                 \\
GNN-LSTM                             & 1                        & 30                  & 0.0034                 \\ \bottomrule
\end{tabular}
\end{table}

\subsubsection{Sliding window deployment}
This section focuses on the deployment of $\calD$ in a sliding window setting. In this experiment, the goal is to see if the proposed FDIA detection can correctly determine whether the inference step $t$ is under attack. Specifically, given an adversarial time window $\calW \equiv \{t-5,...,t\}$, the inference step $t$ is always perturbed, i.e., $\calA(t)=1$. In addition to perturbing the inference timestep $t$, we also consider the possibility of perturbing one or more timesteps in the observation window $\{t-N_p,...,t-1\}$. 

Table \ref{tab:detect-w-infer} shows the detection accuracy of each implementation of $g_\theta$ as a function of the number of adversarial timesteps from $t-5$ to $t-1$.  It can be seen that the detection method achieves high accuracy across three different $g_\theta$ models, especially when there are two or fewer adversarial steps in the observation window. The detection model when using a 100-unit LSTM model achieves better accuracy than using the 50-unit one. 
This implies that a more accurate state predictor is more beneficial for the detection method and generally improves the detection accuracy. In addition, the detection accuracy decreases as the number of adversarial timesteps, $m$, in the observation window increases.

\begin{table}[]
\centering
\caption{Detection accuracy as a function of number of adversarial steps $m$ in the observation window $\{t-5,...,t-1\}$. In this sliding window deployment setting, an adversarial window has the inference step $t$ perturbed, i.e., $\calA(t)=1$.}
\label{tab:detect-w-infer}
\begin{tabular}{cccc}
\hline
$m$ & LSTM ($L_u=100$)   & LSTM ($L_u=50$)  & GNN-LSTM  \\ \hline
0                        & 0.9861 & 0.9829 & 0.9859   \\
1                        & 0.9875 & 0.9829 & 0.9861   \\
2                        & 0.9720 & 0.9594 & 0.9645   \\
3                        & 0.9450 & 0.9176 & 0.9200   \\
4                        & 0.8989 & 0.8534  & 0.8850  \\
5 & 0.8467 & 0.8029 & 0.8328 \\\hline
\end{tabular}
\end{table}

\begin{figure}
    \centering
    \includegraphics[width=0.75\linewidth]{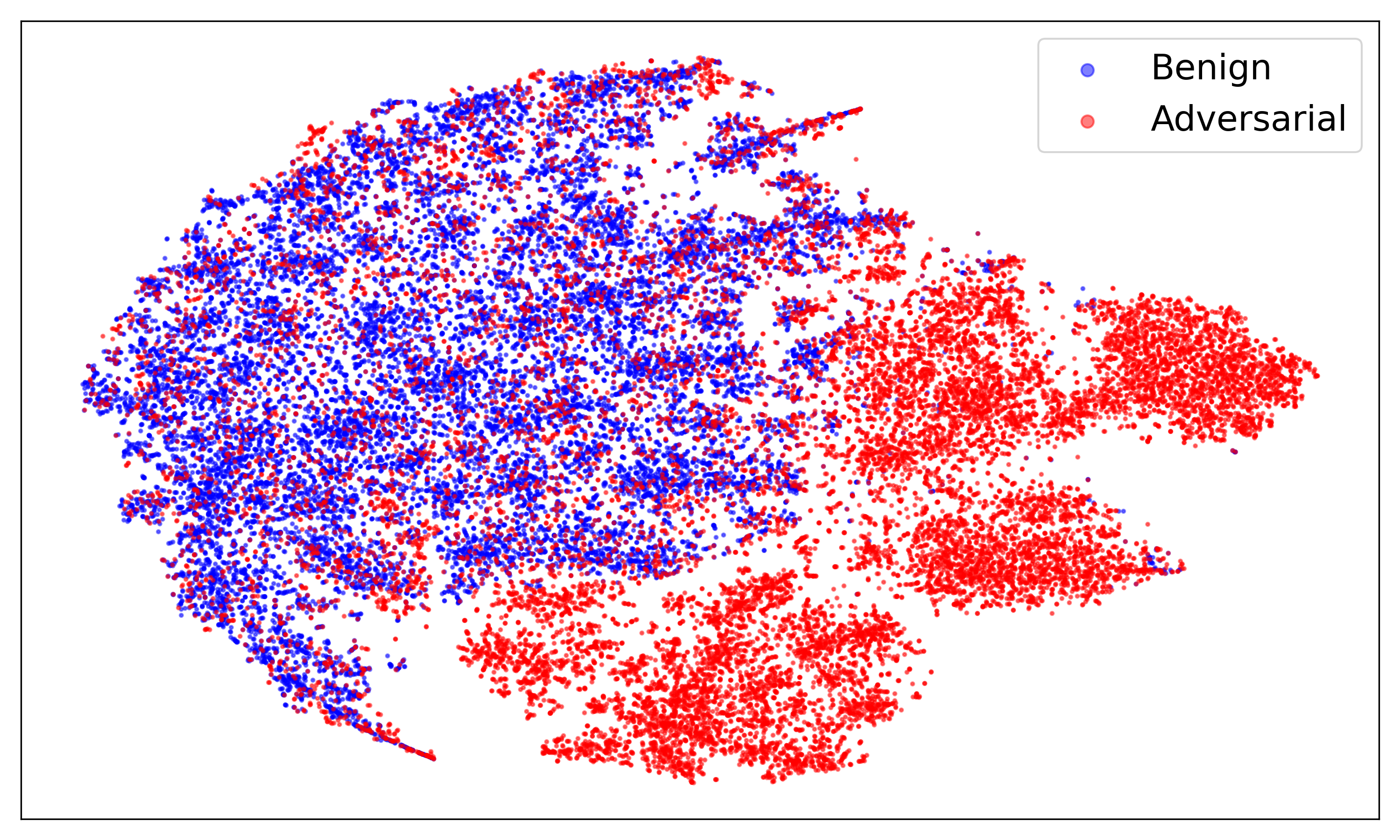}
    \caption{t-SNE visualization of $e_t^\textsf{benign}$ and $e_t^\textsf{adv}$ with 1 adversarial step in the observation window in a sliding window setting}
    \label{fig:tsne-3-w-infer}
\end{figure}

\begin{figure}
    \centering
    \includegraphics[width=0.75\linewidth]{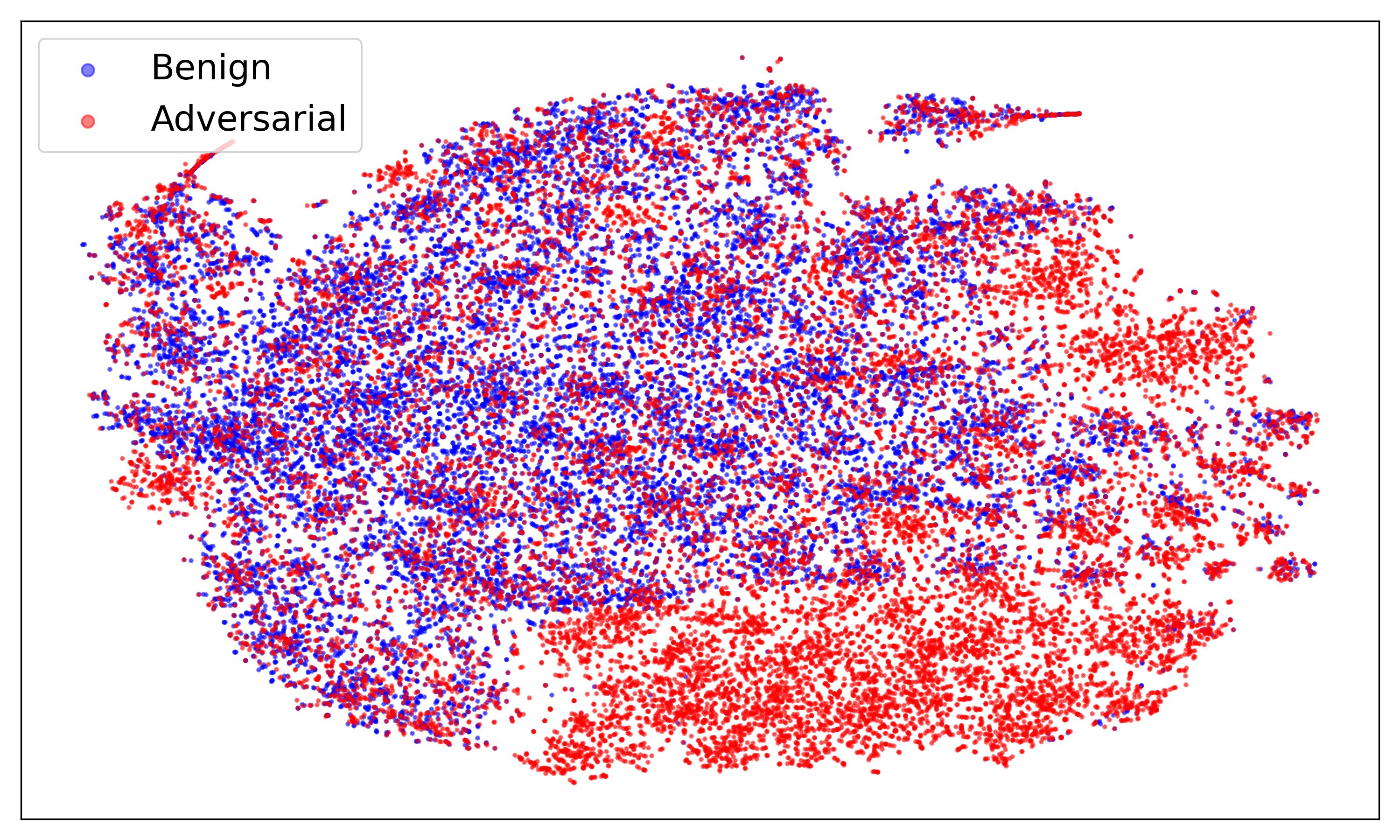}
    \caption{t-SNE visualization of $e_t^\textsf{benign}$ and $e_t^\textsf{adv}$ with 5 adversarial steps in the observation window in a sliding window setting}
    \label{fig:tsne-5-w-infer}
\end{figure}

To give more insights into this result, especially on the impact of $m$ on the detection accuracy, a t-SNE visualization on the input of $h_\phi$, i.e., the prediction error $e_t$, is shown below. The t-distributed Stochastic Neighbor Embedding (t-SNE) representation is a well-known technique to visualize high-dimensional data based on a nonlinear dimensionality reduction mechanism that gives each data point a location in a 2D or 3D space \cite{van2008visualizing}. Figs. \ref{fig:tsne-3-w-infer} and \ref{fig:tsne-5-w-infer} show the t-SNE plots of the prediction error $e_t$ of the 100-unit LSTM model with benign and adversarial windows in the test data when $m=1$ and $m=5$, respectively. Let $e_t^\textsf{benign}$ and $e_t^\textsf{adv}$ denote the prediction error that corresponds to a benign window and an adversarial window, respectively. These plots show that there is a higher distinction between the distribution of $e_t^\textsf{adv}$ and $e_t^\textsf{benign}$ when $m=1$, making it easy for $h_\phi$ to distinguish between the benign and adversarial windows, hence the high accuracy.

If an attacker persists in attacking the system multiple times (high $m$) one would expect a higher effect on the system \cite{rlfdi_prev_paper} and therefore an easier detection. However, the opposite trend is systematically observed in Table~\ref{tab:detect-w-infer}. To explain this phenomenon, recall that, as stated in Section \ref{ssec:detect}, the detector tries to determine whether $x_t$ is abnormal with respect to $x_{t-5}...x_{t-1}$ via the prediction error (note that the $x_t$ is always perturbed in this sliding window setting). Thus, in the case of $m=0$, i.e., no attack occurs in the observation window, and the predicted state $\hat{x}_t$ represents an expected normal measurement at timestep $t$. As a result, there is a high discrepancy between $\hat{x}_t$ and the adversarial measurement $x_t$. On the other hand, when all timesteps in the observation window are attacked, i.e., $m=5$, the prediction of $g_\theta$ is poisoned by the input adversarial observations. This induces an abnormal predicted state  $\hat{x}_t$, thereby making the difference between $\hat{x}_t$ and $x_t$ less distinctive, as demonstrated in Fig. \ref{fig:tsne-5-w-infer}.

\subsubsection{Cyclic Deployment}
\label{cyclic_deployment}
As mentioned in Section \ref{sec:introduction}, a dynamical system has a relatively short interval between timesteps, hence, in the case where $\calD$ cannot be deployed at every single timestep (e.g., due to computational resource constraints), an attack cannot be always assumed to perturb the inference step $t$. For a worst-case analysis, $\calD$ is considered to be deployed periodically every 5 timesteps and the attack only happens within the observation window (i.e., there are one or more adversarial timesteps within $t-5$ to $t-1$), and not at the inference step. In other words, given an adversarial window $\calW \equiv \{t-5,...,t\}$, $\calA(t) = 0$ and $\exists j \in \{t-5,...,t-1\}: \calA(j) = 1$. This presents a challenge for the detector as the current timestep $t$ is not under attack, but a decision must still be made to determine whether an attack occurs in the observation window by using only knowledge of the prediction error $e_t$.

\begin{figure}
    \centering
    \includegraphics[width=0.75\linewidth]{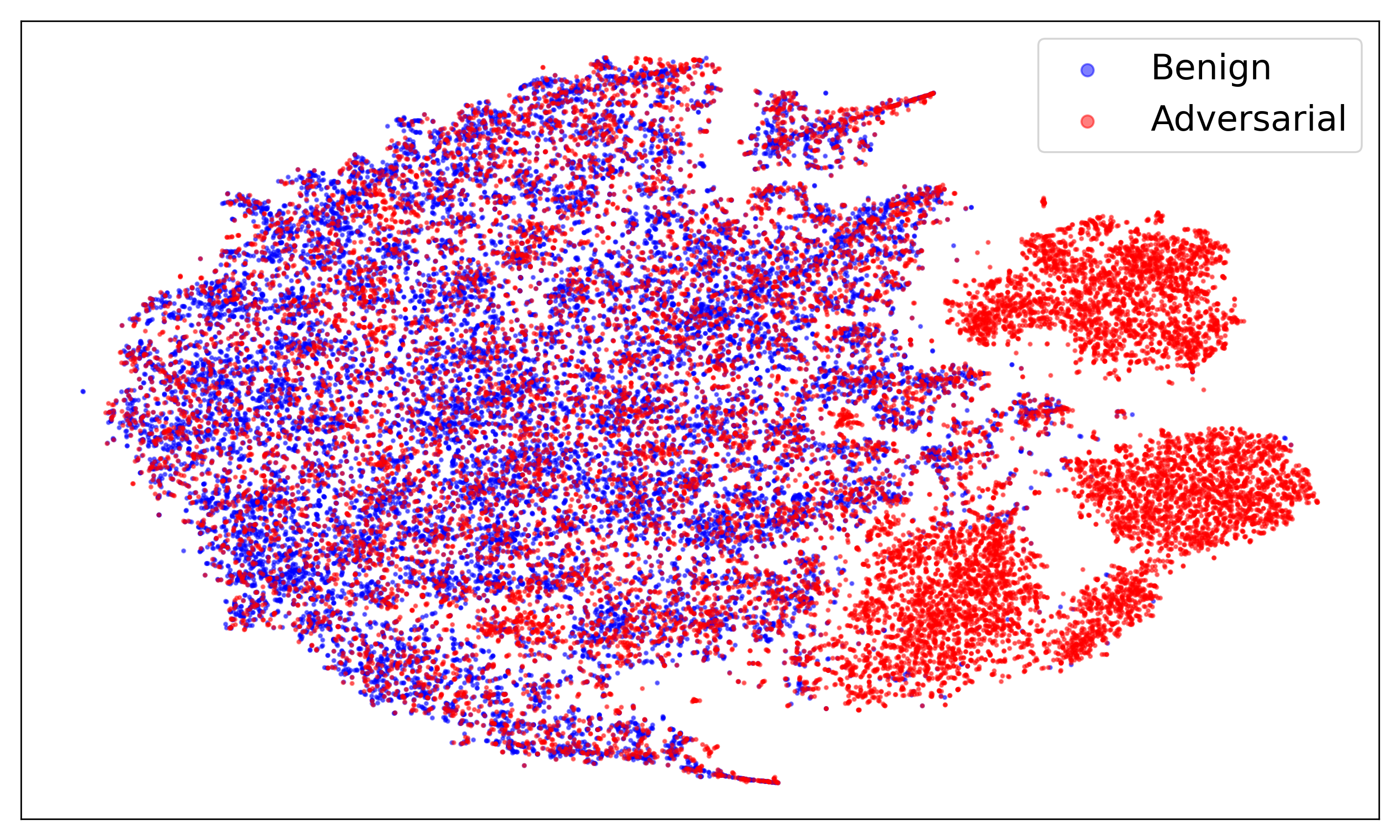}
    \caption{t-SNE visualization of $e_t^\textsf{benign}$ and $e_t^\textsf{adv}$ with 1 adversarial step in the observation window in a cyclic setting}
    \label{fig:tsne-4-no-infer}
\end{figure}

\begin{figure}
    \centering
    \includegraphics[width=0.75\linewidth]{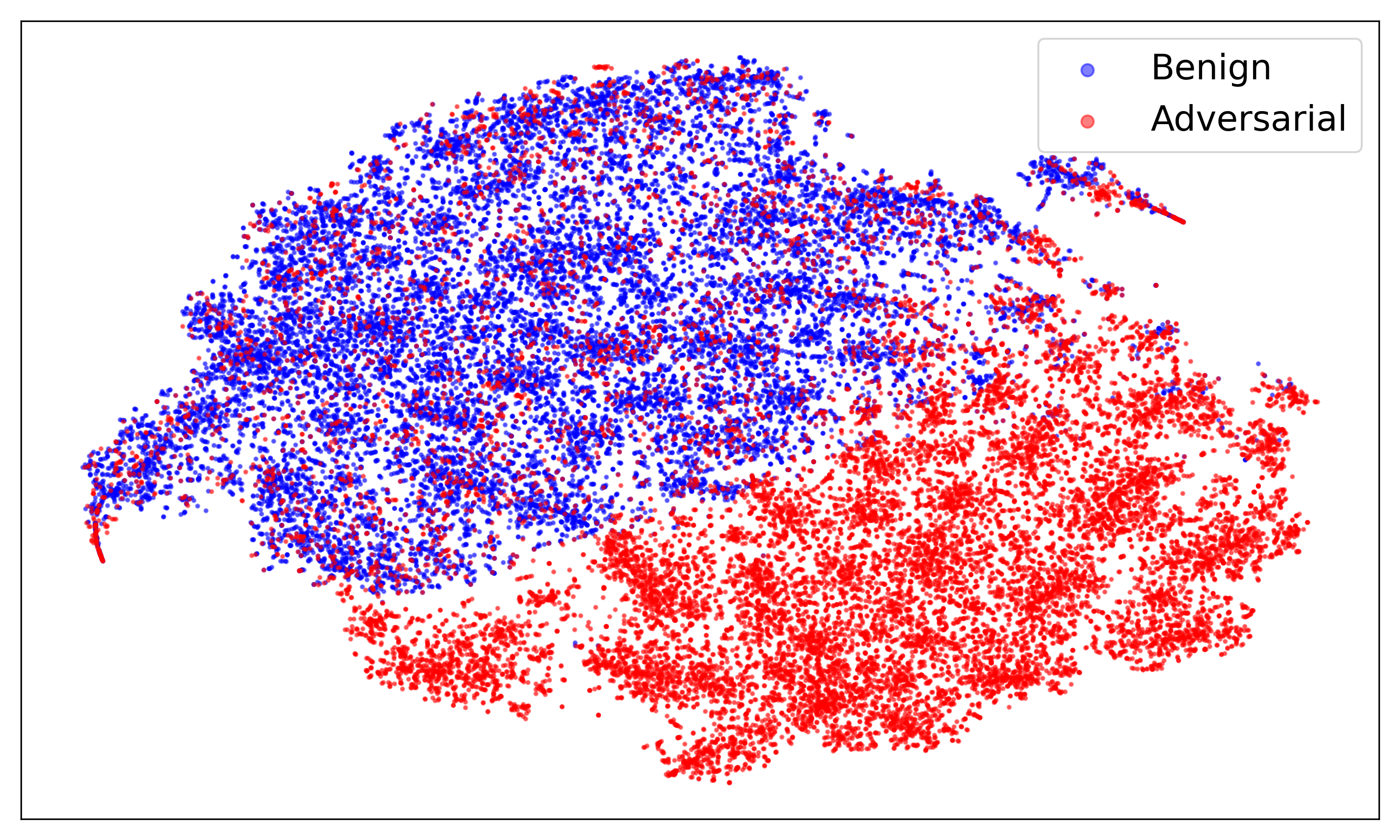}
    \caption{t-SNE visualization of $e_t^\textsf{benign}$ and $e_t^\textsf{adv}$ with 5 adversarial steps in the observation window in a cyclic setting}
    \label{fig:tsne-5-no-infer}
\end{figure}

Table \ref{tab:detect-wo-infer} illustrates the detection accuracy as a function of the number of adversarial timesteps from $t-1$ to $t-5$. Similar to Table \ref{tab:detect-w-infer}, the detection model also attains better accuracy when using a 100-unit LSTM model compared to using a 50-unit.
For this experimental setting, in contrast, the result exhibits a reversed behavior compared to the previous setting where the detection excels when there are more adversarial timesteps in the observation window. Figs. \ref{fig:tsne-4-no-infer} and \ref{fig:tsne-5-no-infer} illustrate the t-SNE plots of the prediction errors with adversarial and benign windows for the cases of 1 and 5 adversarial steps, respectively. It can be seen that the prediction errors between benign and adversarial windows are more distinctive in the latter case.

\begin{table}[]
\caption{Detection rate as a function of the number of adversarial steps $m$ in the observation window $\{t-5,...,t-1\}$. In this setting, $\calA(t) = 0$}
\label{tab:detect-wo-infer}
\centering
\begin{tabular}{cccc}
\hline
$m$ & LSTM ($L_u=100$)   & LSTM ($L_u=50$)  & GNN-LSTM  \\ \hline
1                        & 0.7869 & 0.7653 & 0.7655   \\
2                        & 0.9288 & 0.9143  & 0.8978  \\
3                        & 0.9837 & 0.9752  & 0.9800  \\
4                        & 0.9911 & 0.9886  & 0.9891  \\
5 & 0.9917 & 0.9891 & 0.9891 \\\hline
\end{tabular}
\end{table}

This behavior can be explained by the fact that if the state predictor $g_\theta$ is heavily poisoned by the adversarial observation window (as in the case of $m=5$), the predicted state $\hat{x}_t$ becomes abnormal, thereby increasing the prediction error between $\hat{x}_t$ and $x_t$ (note that in this setting, there is no perturbation at $t$). On the other hand, when $m=1$, the prediction is less influenced by the adversarial measurements, making the difference between $\hat{x}_t$ and $x_t$ less distinctive as demonstrated in Fig. \ref{fig:tsne-4-no-infer}. 

\begin{figure}
    \centering
    \includegraphics[width=0.75\linewidth]{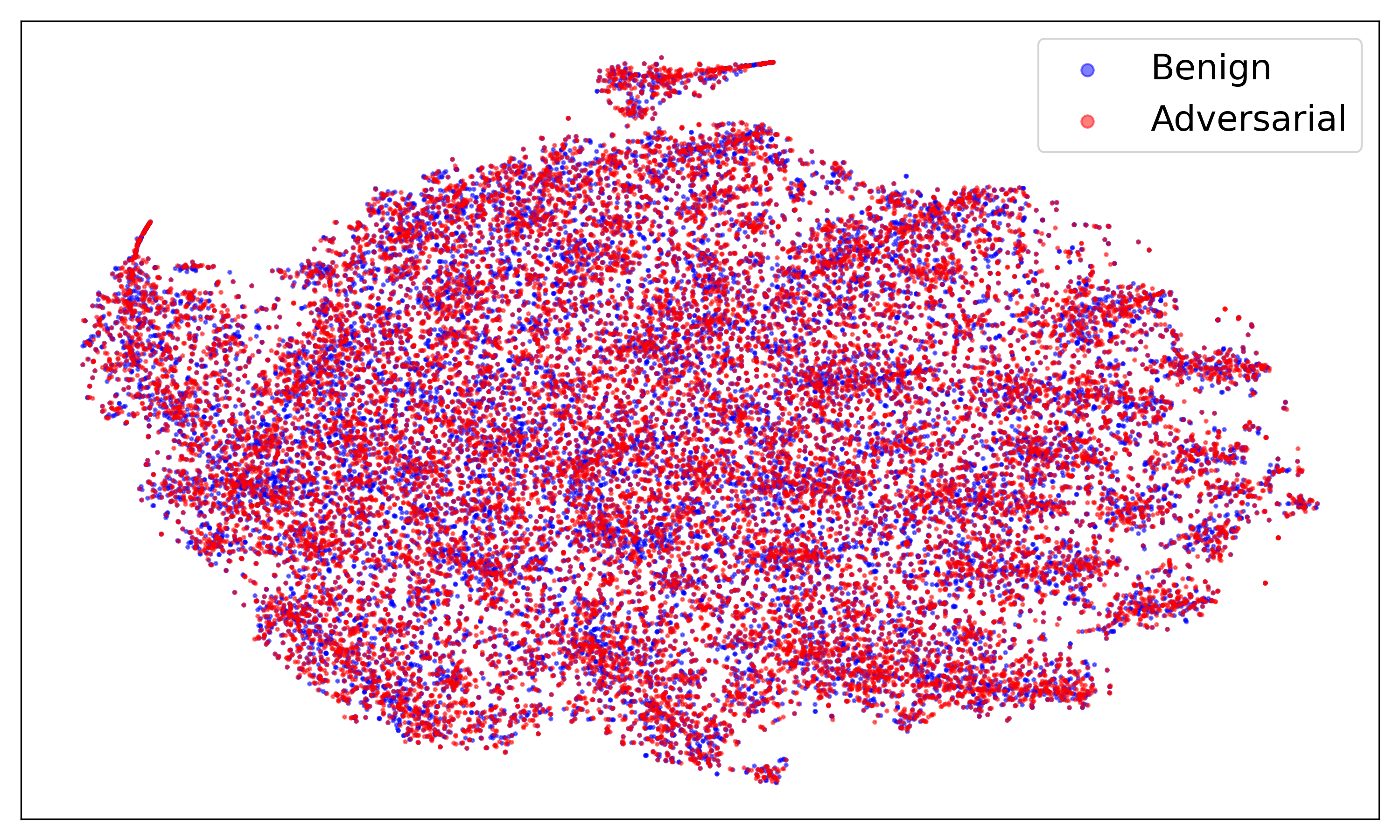}
    \caption{t-SNE visualization with one adversarial timestep at $t-5$}
    \label{fig:tsne-1st}
\end{figure}

Table \ref{tab:detect-pos} shows how the position of the adversarial timestep affects the detection accuracy with $g_\theta$ being a 100-unit LSTM. For simplicity, only the case where there is one adversarial timestep in the observation window is considered. It can be seen that the detection becomes less effective as the adversarial timestep is placed farther away from the inference step $t$. This phenomenon suggests that any perturbation to timesteps that are farther away from the inference step has less effect on the prediction error. This is in line with the characteristics of an LSTM model. 
In fact, Fig.  \ref{fig:tsne-1st} demonstrates that, when the adversarial timestep is at $t-5$, the prediction errors of benign and adversarial windows are indistinguishable.


\begin{table}
\centering
\caption{Detection rate as a function of positions of the adversarial timestep. The state predictor $g_\theta$ is an LSTM model with $L_u = 100$}
\label{tab:detect-pos}
\begin{tabular}{clllll}
\toprule
Adversarial position & Accuracy & F1 Score & Precision & Recall \\
\midrule
$t-1$ & 0.9888 & 0.9887 & 0.9992 & 0.9784 \\
$t-2$ & 0.9818 & 0.9816 & 0.9962 & 0.9673 \\
$t-3$ & 0.9261 & 0.9221 & 0.9751 & 0.8745 \\
$t-4$ & 0.6593 & 0.6366 & 0.6822 & 0.5968 \\
$t-5$ & 0.5115 & 0.4333 & 0.5161 & 0.3734 \\
\bottomrule
\end{tabular}
\end{table}

\subsubsection{Detection rate with noisy measurements}

\begin{figure*}[ht!]
    \centering
  \subfigure[LSTM w/ Sliding Window]{\includegraphics[width=0.49\columnwidth]{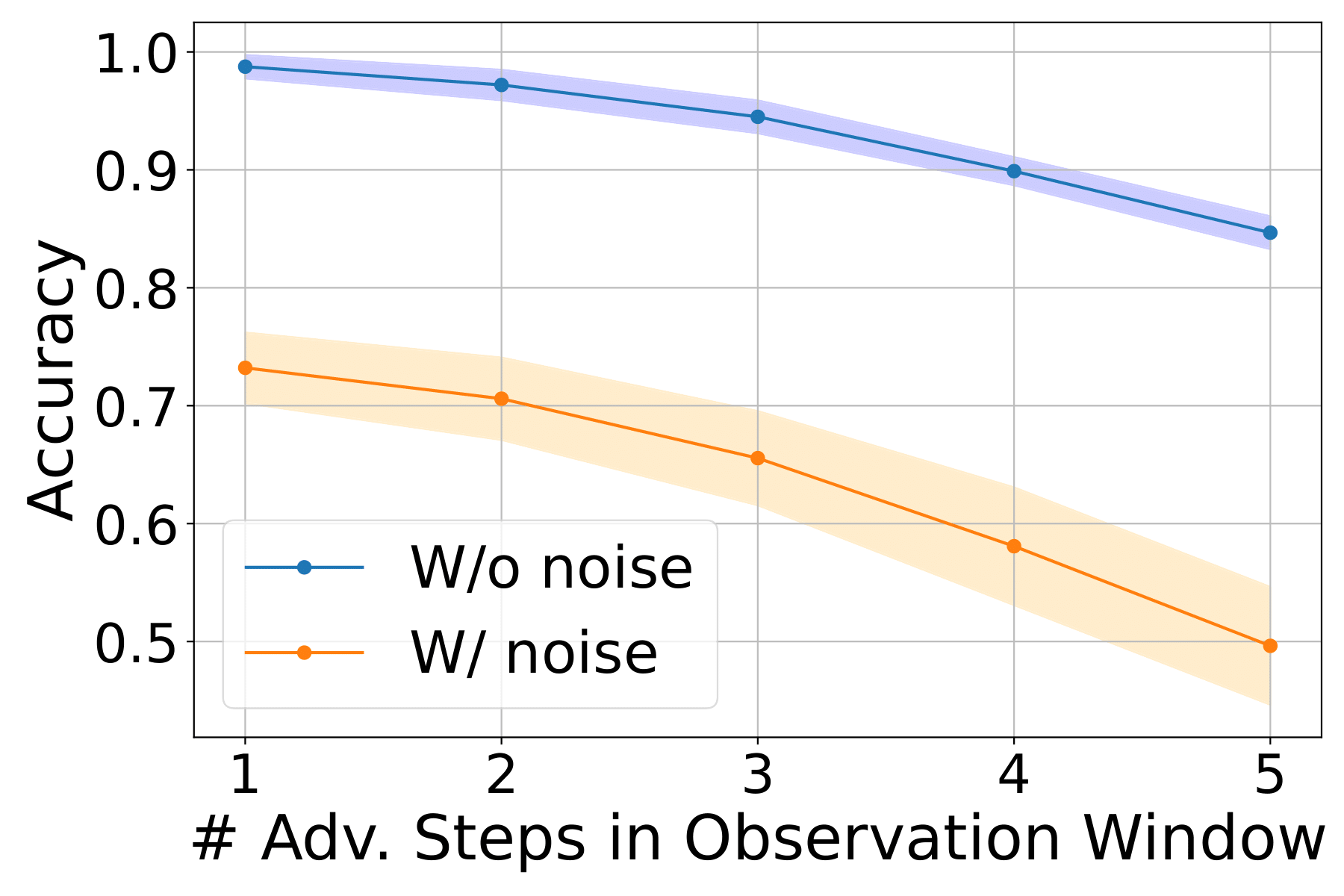}}%
      \subfigure[LSTM w/ Cyclic]{\includegraphics[width=0.49\columnwidth]{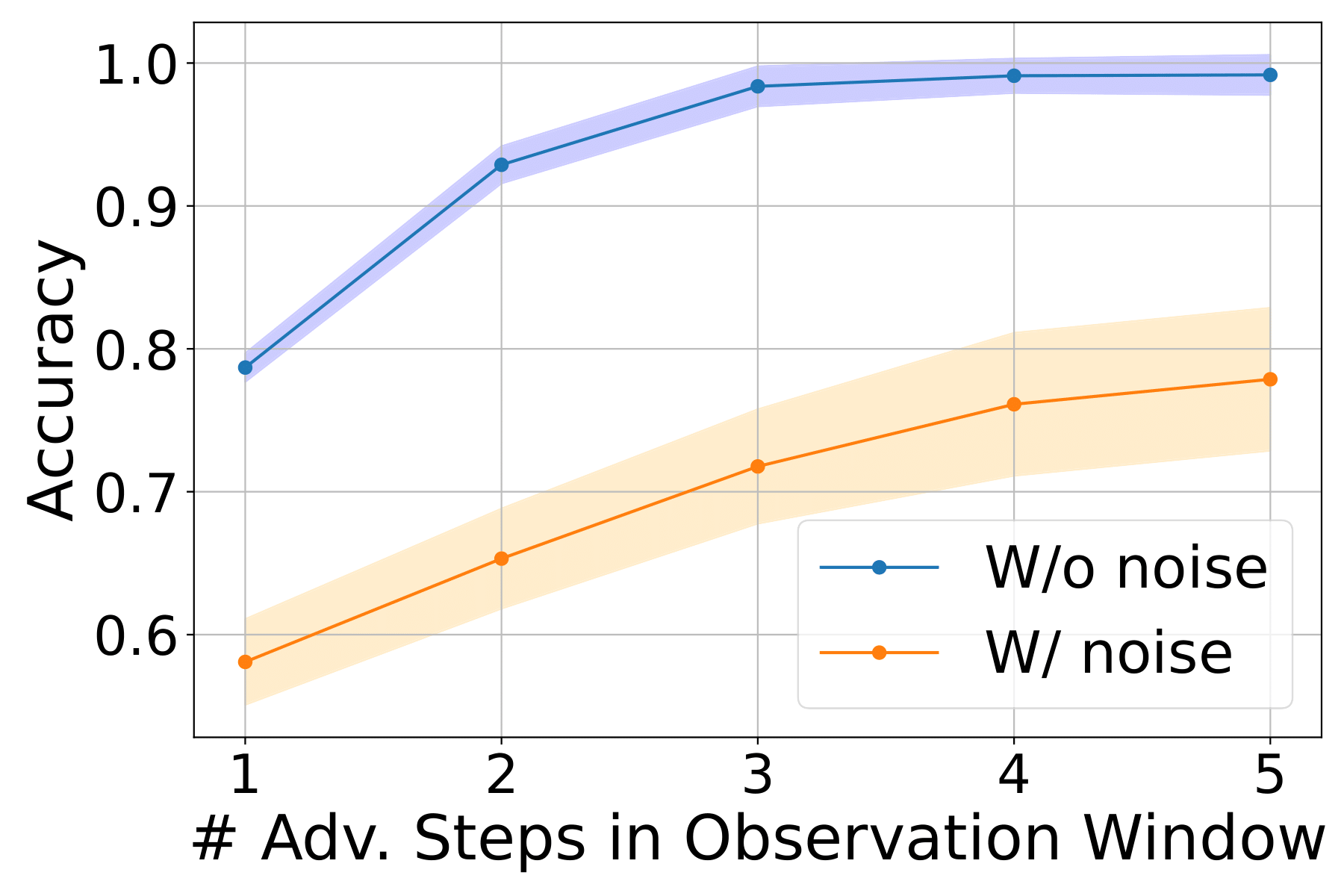}}%
    \subfigure[GNN-LSTM w/ Sliding Window]{\includegraphics[width=0.49\columnwidth]{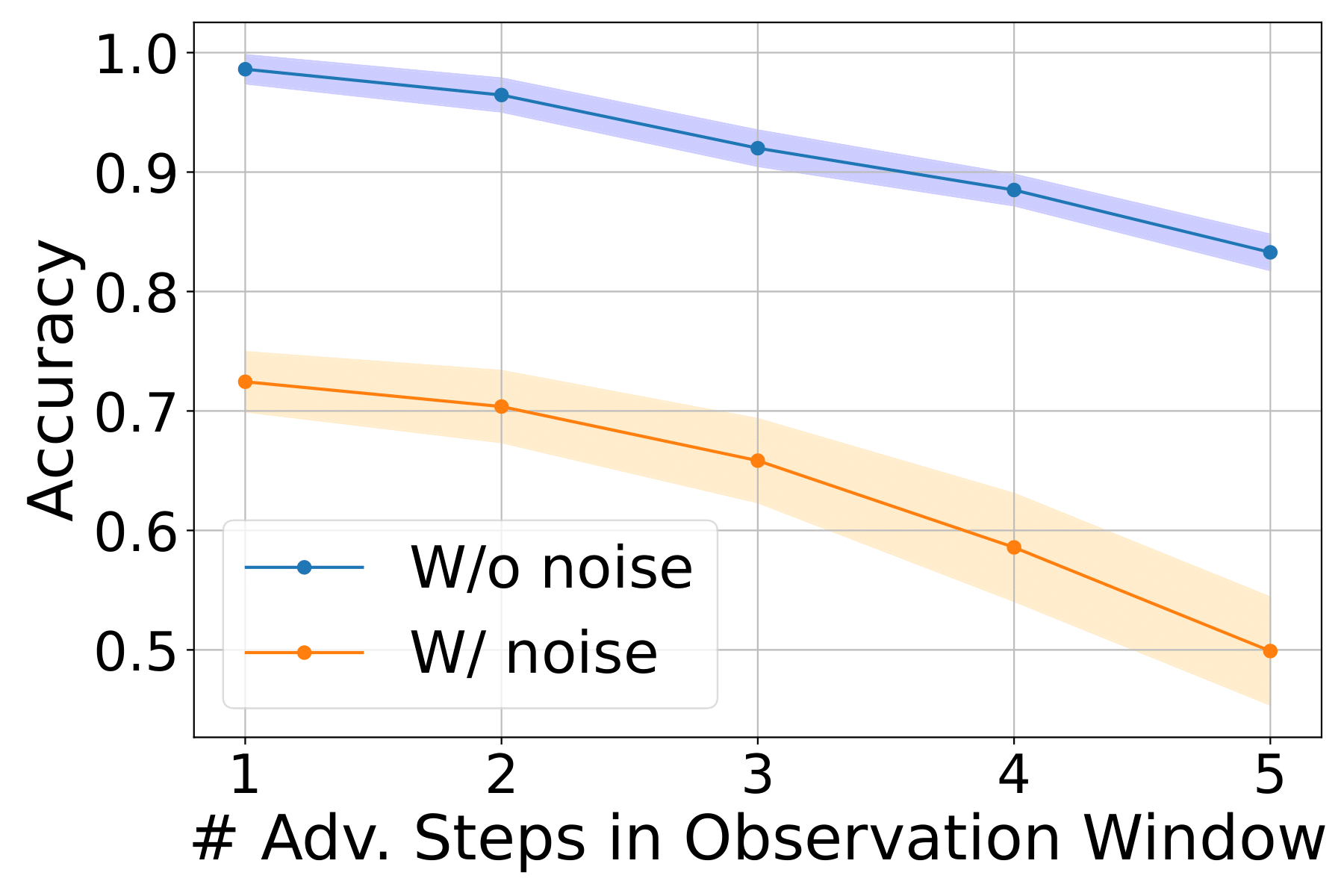}}%
    \subfigure[GNN-LSTM w/ Cyclic]{\includegraphics[width=0.49\columnwidth]{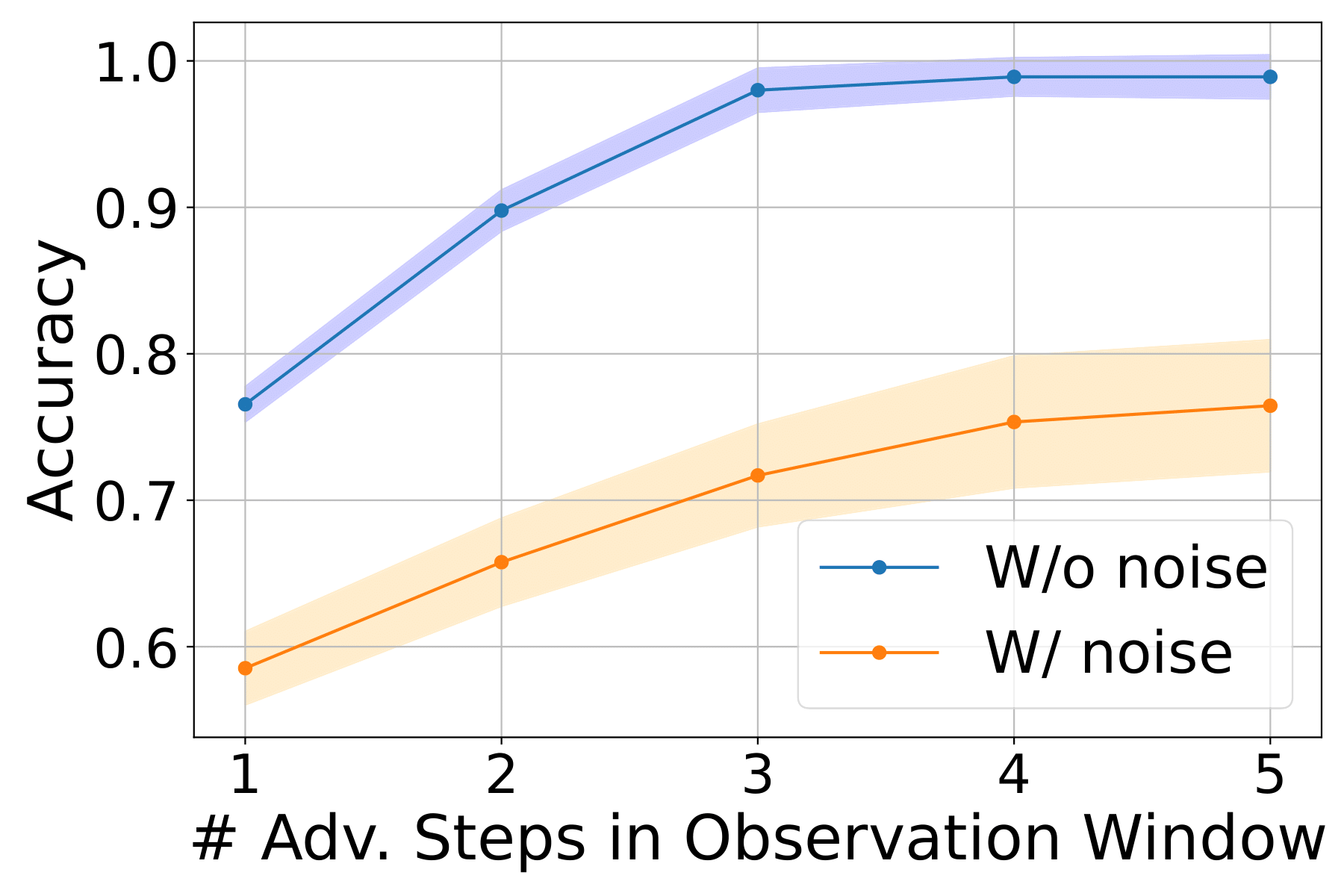}}%
    \caption{Detection accuracy of $\calD$ under noise with two state prediction models: 100-unit LSTM and GNN-LSTM. Each model is tested under two deployment settings. The randomness is over the training algorithm of $h_\phi$ and the noise added to the observations.}
    \label{fig:detect-noise}
\end{figure*}

In this section, the objective is to characterize the efficacy of the detection method $\calD$ in the presence of noise. The LSTM (with $L_u=100$) and GNN-LSTM models that were trained with noise (Section \ref{lstm_noise}) are used as $g_\theta$, and then the classifier $h_\phi$ is trained with noise added to the dataset. In this experiment, the noise level is $\sigma=0.001$, i.e. an additive noise distributed as $\mathbb{N}(0,0.001)$ is added to every state variable.

Fig. \ref{fig:detect-noise} shows the detection accuracy with LSTM and GNN-LSTM as $g_\theta$ under noise. It can be seen that, compared to the results on clean data, the noisy observations worsen the detection accuracy in both deployment settings. Note that according to Section \ref{sec:lstm_gnn}, the MAE of the state prediction method increases by 9 times under noise. This result further reinforces the implication that the performance of state predictor $g_\theta$ influences the detection accuracy. Furthermore, Fig.~\ref{fig:detect-noise} shows that the margin of error is higher under noisy conditions, suggesting that the detection model is more uncertain when encountering with noisy observations.

\subsubsection{Localizing FDIA via 
Multiclass Classifier}\label{sec:multibus}
This experiment extends our detection method to localizing an FDIA, that is, pinpointing which bus is under attack. To achieve this, $h_\phi$ is transformed into a multiclass classifier that maps the prediction error from $g_\theta$ to an output label indicating the bus index that is perturbed by the FDIA, i.e., $h_\phi : \mathbb{R}^d \rightarrow \{0,1,...,9, \bot\}$ (where $\bot$ denotes no attack). For this experiment, $h_\phi$ has 1 hidden layer with 100 neurons and is trained using an Adam optimizer with a learning rate of 0.0005. 


\begin{table}[h]
\centering
 \caption{Detection rate of different buses under five adversarial timesteps in the observation window}
\begin{tabular}{cccc} 
 \hline
 Bus No. & F1-score & Precision  & Recall   \\
 \hline
0 &  0.89  &  0.82   &   0.97        \\ 
1 & 0.93 & 0.90   &    0.97       \\
2 & 0.94   & 0.92   &   0.96         \\
3 & 0.91 & 0.84   &   0.99         \\
4 &  0.95 & 0.93   &   0.99        \\
5 &  0.92  & 0.88   &   0.97      \\
6 &  0.80 & 0.79    &  0.81       \\
7 &  0.96 & 0.94    &  0.99      \\
8 &   0.91 &  0.86    &  0.95       \\
9 &     0.41  &  0.60   &    0.31       \\
 \hline
 \end{tabular}
\label{tab:detect-wo-infer_multilabel_ad5}
\end{table}

Table \ref{tab:detect-wo-infer_multilabel_ad5} illustrates the detection accuracy per bus index. With the exception of bus 9, the detection model is able to localize FDIAs with high accuracy, especially if the attack occurs at bus 7. This behavior, in fact, corroborates with the findings in previous work \cite{rlfdi_prev_paper} which shows that perturbing bus 9 induces \textit{negligible} frequency oscillations while perturbing bus 7 has the highest impact. Intuitively speaking, an attack with a higher impact would be more detectable. The overall accuracy across all buses is 0.84, thus demonstrating that the proposed FDIA detection method based on state prediction not only can detect attacks with high accuracy but can also be easily extended to effectively localize them.



\subsection{Discussion}
\label{sec:eff_disc}
From the experimental results, it can be inferred that the proposed state prediction method can be used as a building block for developing an FDIA detection that attains good performance across various attack and deployment settings. In particular, these experiments have shown that the detection method does not need to be deployed at every single timestep to remain effective. Instead, the cyclic deployment setting enables a low-cost and efficient way to execute the detection model. The results also suggest that the detection accuracy is heavily influenced by the predictive performances of the state prediction method, and that the proposed detection can be easily extended, if needed, to localize FDIAs.

Nonetheless, the results also signify some potential vulnerabilities of this detection method can be exploited by an attacker. There are two attack strategies that could evade detection. First, since the detection method is a composition of $g_\theta$ and $h_\phi$, one strategy is to attack both models at the same time, meaning an attacker can perturb the observation window to poison the $g_\theta$ to give out an incorrect state prediction, and then perturb the inference step to manipulate the prediction error. However, as shown in Table \ref{tab:detect-w-infer}, this strategy is not very effective because the model can still maintain at least 0.84 accuracy even when the attacker perturbs the inference step and all the steps in the observation window. The second strategy is to tactically perturb only one timestep in the observation window that is far away from the inference step, e.g., at $t-4$ or $t-5$ w.r.t. an observation window size of 5. However, this strategy requires that an attacker knows specifically at which timestep the detection model is deployed. Such deployment information might not be easily accessible to attackers. Moreover, this attack strategy can be countered by dynamically changing the deployment interval (in a cyclic setting) over time.

Finally, from Table \ref{tab:detect-w-infer}, \ref{tab:detect-wo-infer}, and \ref{tab:detect-pos}, it is recommended that, for the cyclic deployment, the frequency at which the detection model is deployed should be every 4 timesteps (resulting in an observation window size $N_p=3$). This is because, in the worst-case scenario where an adversary perturbs one timestep at $t-3$, the detection model still maintains an accuracy greater than 0.92. This setting should, however, be weighed against the cost of deploying detection more frequently.



\section{Conclusions}

 This paper presented a state-prediction-based intrusion detection system for a power dynamical system using a temporal LSTM and a spatio-temporal GNN-LSTM for state-prediction. The state predictors are paired with a single or multi-class classifier, allowing to detect FDIA with both sliding window and cyclical settings. The performance of the LSTM outperformed GNN-LSTM predictor in the presence of noise. Finally, the FDIA detection method was shown to attain high accuracy under two deployment settings: sliding window and cyclic. This shows that a data-based state prediction mechanism can be used to build a reliable and computationally efficient FDIA detection model for power system dynamics.

 \section*{Acknowledgment}
This work was authored by the National Renewable Energy Laboratory (NREL), operated by Alliance for Sustainable Energy, LLC, for the U.S. Department of Energy (DOE) under Contract No. DE-AC36-08GO28308. This work was supported by the Laboratory Directed Research and Development (LDRD) Program at NREL. The views expressed in the article do not necessarily represent the views of the DOE or the U.S. Government. The U.S. Government retains and the publisher, by accepting the article for publication, acknowledges that the U.S. Government retains a nonexclusive, paid-up, irrevocable, worldwide license to publish or reproduce the published form of this work, or allow others to do so, for U.S. Government purposes. This research was performed using computational resources sponsored by the Department of Energy's Office of Energy Efficiency and Renewable Energy and located at the National Renewable Energy Laboratory.

\bibliographystyle{IEEEtran}

\newpage

\begin{IEEEbiography}[{\includegraphics[width=1in,height=1.25in,clip,keepaspectratio]{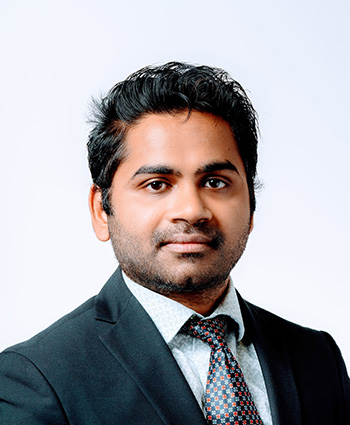}}]{Abhijeet Sahu} (Member, IEEE) received the M.S and Ph.D. degrees in electrical engineering and computer engineering (ECEN) from Texas A\&M University at College Station, TX in 2018 and 2022 respectively.  He holds a B.S. degree in electronics and communications from National Institute of Technology, Rourkela, India, in 2011. He is currently a Senior Research Engineer working in cybersecurity at the National Renewable Energy Laboratory (NREL). His research interests include network security, cyber-physical modeling for intrusion detection and response, and artificial intelligence for cyber-physical security in power systems.
\end{IEEEbiography}

\begin{IEEEbiography}[{\includegraphics[width=1in,height=1.25in,clip,keepaspectratio]{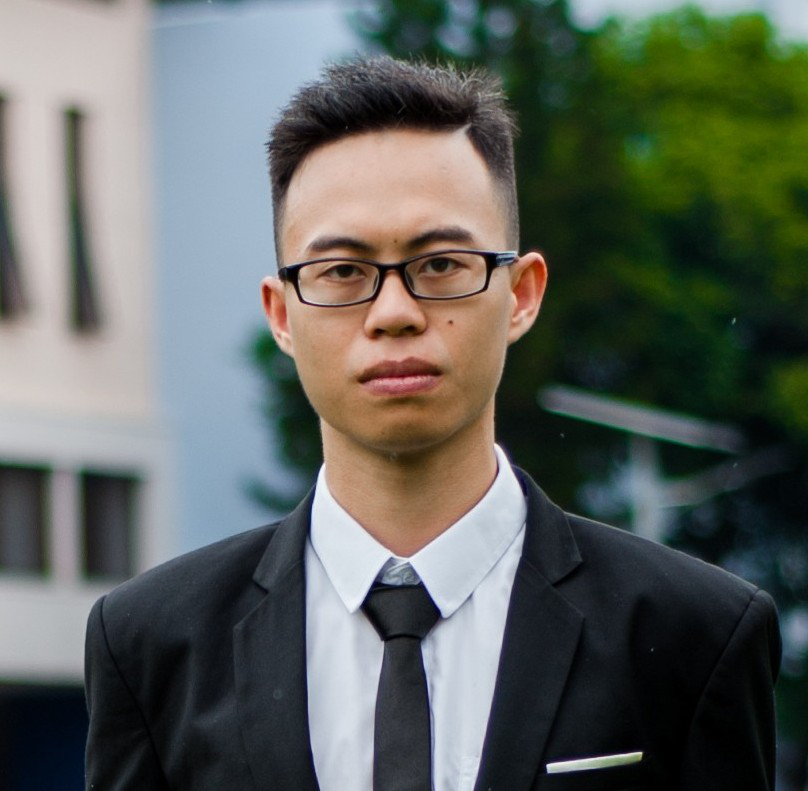}}]{Truc Nguyen} received his Ph.D. degree in computer engineering from the University of Florida in 2023. Before that, he received his Bachelor of Engineering degree in computer engineering from Ho Chi Minh City University of Technology in 2018. He is currently working at the National Renewable Energy Laboratory (NREL). His research interests include trustworthy machine learning and cybersecurity.
\end{IEEEbiography}

\begin{IEEEbiography}[{\includegraphics[width=1in,height=1.25in,clip,keepaspectratio]{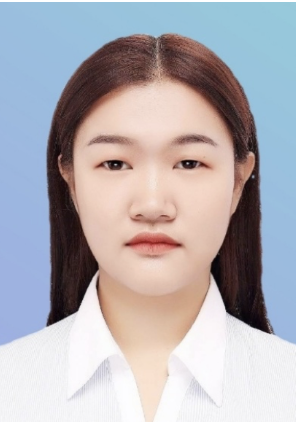}}]{Kejun Chen} is currently pursuing a Ph.D. degree in the ECE Department of UC Santa Cruz. She is currently a graduate student intern at the National Renewable Energy Laboratory. Her research interests focus on the application of machine learning in power system operation. 
\end{IEEEbiography}

\begin{IEEEbiography}[{\includegraphics[width=1in,height=1.25in,clip,keepaspectratio]{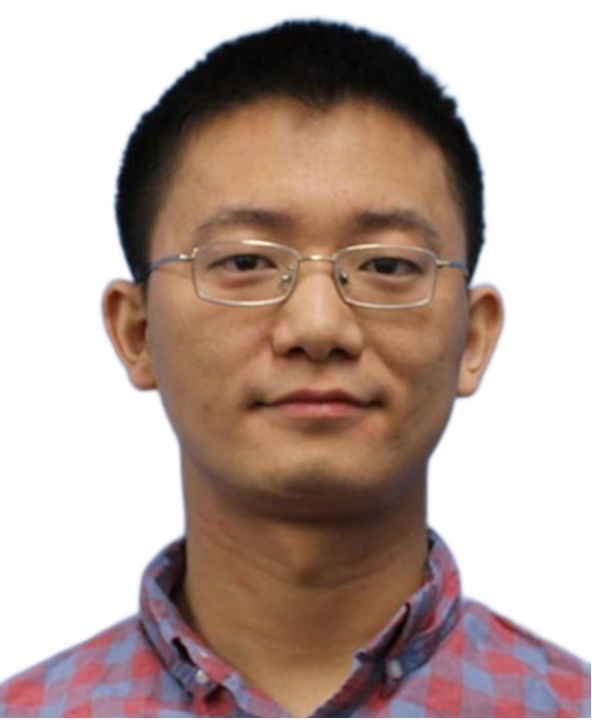}}]{Xiangyu Zhang} received his Ph.D. degree in Electrical Engineering from Virginia Tech. He was a researcher in the Computational Science Center of NREL when this research work is conducted. His research interests mainly cover learning-based optimal control of energy systems.
\end{IEEEbiography}

\begin{IEEEbiography}[{\includegraphics[width=1in,height=1.25in,clip,keepaspectratio]{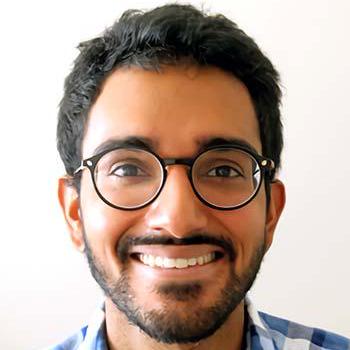}}]{Malik Hassanaly}
earned a Ph.D. in Aerospace Engineering from the University of Michigan and is currently a researcher in the computational science center of NREL. His research focuses on uncertainty quantification and scientific machine learning to accelerate or augment physics-based models.
\end{IEEEbiography}

\clearpage
\appendices
\onecolumn
\section{Hyperparameter tuning for the FDIA classifier}
\label{sec:app-tune}
This section outlines the hyperparameter tuning process for the classifier $h_\phi$ when conducting experiments in Section \ref{detection_result}. As above-mentioned, with respect to each implementation of the state predictor $g_\theta$, Optuna \cite{akiba2019optuna} is used to tune hyperparameters of $h_\phi$ via Tree-structured Parzen Estimator algorithm. With $h_\phi$ being an MLP, the following hyperparameters and search space are considered:

\begin{table}[h!]
\centering
\begin{tabular}{@{}ll@{}}
\toprule
\textbf{Hyperparameters}               & \textbf{Search Space} \\ \midrule
Number of hidden layers (n\_layers)               & [1, 3]             \\
Number of neurons in each hidden layer (n\_units) & [10, 150]         \\
Learning rate (lr)                         & [1e-4, 1e-1]      \\ \bottomrule
\end{tabular}
\end{table}

The model is tuned over 100 trials and Optuna returns the best hyperparameters in the search space that result in the minimum loss. The plot below illustrates the optimization history over 100 trials, the Objective Value is the loss on the validation set.

\begin{figure}[h!]
    \centering
    \includegraphics[width=\linewidth]{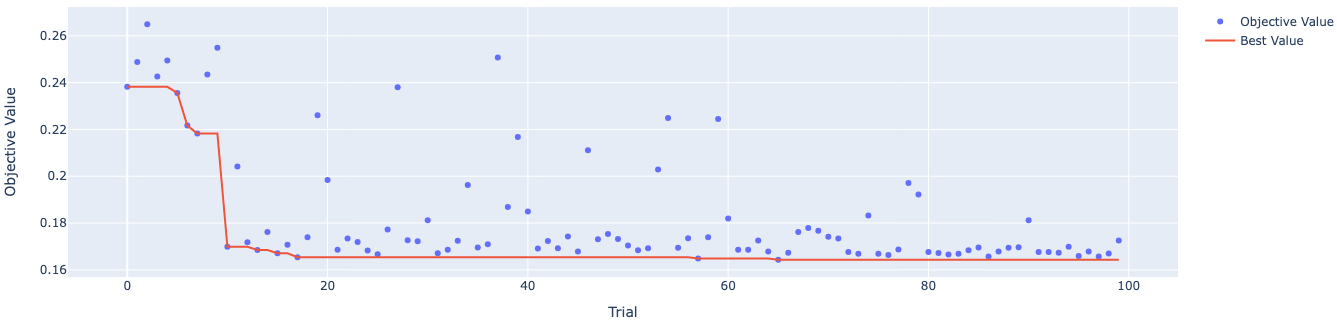}
\end{figure}

The next figure shows the importance score of each hyperparameter. It can be seen that the number of hidden layers (n\_layers) is the most impactful one, while the learning rate (lr) is the least.

\begin{figure}[h!]
    \centering
    \includegraphics[width=\linewidth]{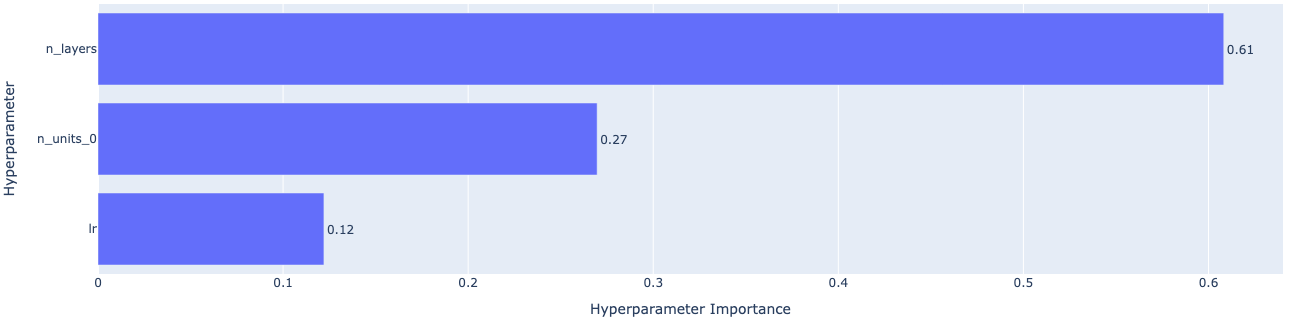}
\end{figure}

Upon inspecting the tuning process, we have seen that setting the number of hidden layers to 1 results in the best model performance. With 1 hidden layer, the figure below shows how the number of neurons (n\_units\_0) in that layer and the learning rate (lr) influence the validation loss (Objective Value) over 100 trials (each black dot represents one trial).

\begin{figure}[h!]
    \centering
    \includegraphics[width=\linewidth]{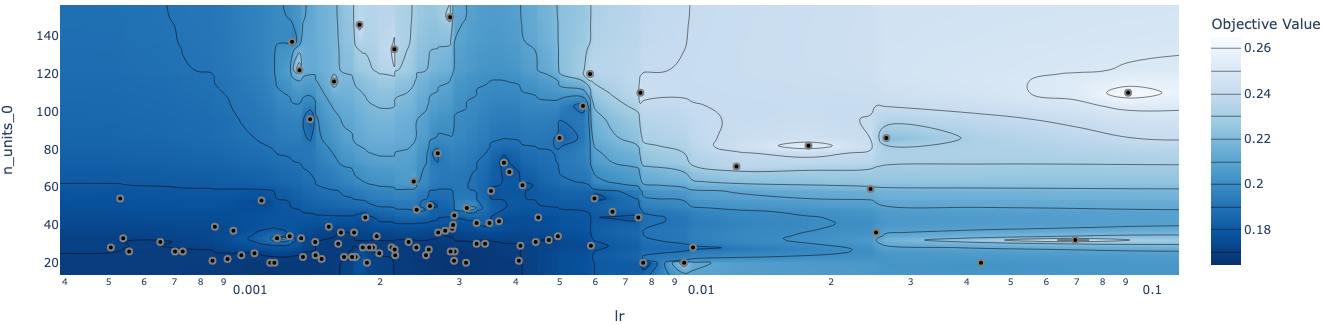}
\end{figure}

The resulting best set of hyperparameters for $h_\phi$ was shown in Table \ref{tab:param} in the main manuscript.

\end{document}